\documentclass[9pt,shortpaper,twoside,web]{ieeecolor2}
\usepackage{generic}
\usepackage{cite}
\usepackage{amsmath,amssymb,amsfonts}
\usepackage{algorithmic}
\usepackage{graphicx}
\usepackage{textcomp}
\usepackage{subfigure}
\usepackage{subcaption}
\usepackage{MnSymbol}
\def\BibTeX{{\rm B\kern-.05em{\sc i\kern-.025em b}\kern-.08em
    T\kern-.1667em\lower.7ex\hbox{E}\kern-.125emX}}
\begin{document}
\title{Impact of a Lower Limb Exosuit Anchor Points on Energetics and Biomechanics}
\author{Chiara Lambranzi, \IEEEmembership{Graduate Student Member, IEEE}, Giulia Oberti, Christian Di Natali, Darwin G. Caldwell, \IEEEmembership{Fellow, IEEE}, Manuela Galli, \IEEEmembership{Member, IEEE}, Elena De Momi, \IEEEmembership{Senior Member, IEEE}, Jesùs Ortiz
\thanks{Manuscript received 20 February 2025; revised 06 June 2025; accepted 23 July 2025.
This work was supported by the Istituto Italiano di Tecnologia.}
\thanks{C. Lambranzi, G. Oberti, C. Di Natali, D. G. Caldwell and J. Ortiz are with Istituto Italiano di Tecnologia, Department of Advanced Robotics, Via San Quirico 19d, Genoa, Italy}
\thanks{C. Lambranzi, G. Oberti, M. Galli and E. De Momi are with Politecnico di Milano, Department of Electronics, Information and Bioengineering, Milan, Italy.}
\thanks{Corresponding author's email: chiara.lambranzi@iit.it}}

\maketitle

\begin{abstract}
Anchor point placement is a crucial yet often overlooked aspect of exosuit design since it determines how forces interact with the human body. This work analyzes the impact of different anchor point positions on gait kinematics, muscular activation and energetic consumption. A total of six experiments were conducted with 11 subjects wearing the XoSoft exosuit, which assists hip flexion in five configurations. Subjects were instrumented with an IMU-based motion tracking system, EMG sensors, and a mask to measure metabolic consumption. The results show that positioning the knee anchor point on the posterior side while keeping the hip anchor on the anterior part can reduce muscle activation in the hip flexors by up to 10.21\% and metabolic expenditure by up to 18.45\%. Even if the only assisted joint was the hip, all the configurations introduced changes also in the knee and ankle kinematics. Overall, no single configuration was optimal across all subjects, suggesting that a personalized approach is necessary to transmit the assistance forces optimally. These findings emphasize that anchor point position does indeed have a significant impact on exoskeleton effectiveness and efficiency. However, these optimal positions are subject-specific to the exosuit design, and there is a strong need for future work to tailor musculoskeletal models to individual characteristics and validate these results in clinical populations.

\end{abstract}


\section{Introduction}
\label{sec:introduction}
In recent decades, the number and proportion of people 60 years and older in the population has been increasing worldwide. Projections indicate that this demographic is expected to reach approximately 2 billion by 2050 \cite{un2022}.
The growth of the elderly population is leading to an increase in age-related conditions, such as reduced mobility and autonomy, decreased strength, and musculoskeletal function. People with impaired movement are at risk of other associated issues, such as social isolation, physical decline, and psychosocial distress.
Empowering impaired individuals with the tools to adapt their interactions with the environment is crucial in preserving their intrinsic capacities, functional ability, and independence.

Wearable assistive technologies, such as exoskeletons, have emerged as a promising solution to aid impaired movement, in addition to conventional assistive devices. 
Most exoskeletons are designed with rigid structures, where the structural frames and links not only support part of the exoskeleton weight but also transmit high forces and torques. This makes rigid exoskeletons particularly suitable when the patients are severely impaired\cite{Sposito2022}. Soft exoskeletons, or exosuits, are an alternative that features a flexible and light structure. 
They are gaining traction in applications where the subject retains some mobility. Since they lack a rigid frame, exosuits are more portable, comfortable and allow for a more natural movement. However, they also require greater effort from the user to generate motion.

The design of these devices should reduce the user's energy consumption, especially when the target populations are easily fatigued subjects.
Interventions aimed at improving aerobic capacity or reducing the energetic cost of walking may help prevent gait slowing and support mobility in older adults \cite{richardson2015}. Therefore, efforts to delay or prevent mobility loss should be aimed at increasing fitness and minimizing energetic cost \cite{schrack2012}.

The design of exosuits for elderly assistance requires careful consideration of the interactions between the users and robots from multiple perspectives.
While several studies prove that these devices can reduce walking fatigue and the related energetic consumption\cite{Martini2019, DiNatali2023, Kim2022}, there is limited research on how the placement of assistive force anchor points influences user energetics. Moreover, the geometric properties of exosuits are often based on intuition and experience rather than rigorous analysis \cite{grabke2019, xiloyannis2022}.

This study builds upon the work of Lambranzi et al. \cite{lambranzi2025}, which used musculoskeletal simulations to identify optimal anchor points in a quasi-passive exosuit. Our previous work conducted experimental validation on only two configurations, focusing solely on metabolic consumption to compare with simulation results. The findings revealed a discrepancy between the simulation and the experimental results, highlighting how simulation alone does not always provide enough information to determine the optimal solution.

In this study, we expand the analysis to assess anchor point placement's effects on biomechanics and energetics. Using the same experimental data, we investigate not only metabolic consumption but also kinematics and muscle activation across all tested configurations in 11 healthy subjects wearing the XoSoft exosuit.
Energy consumption remains a central variable in our approach, as its optimization can reduce fatigue and enhance user motivation, which are key factors in rehabilitation and the daily use of assistive devices.

Section \ref{sec:state_art} reviews the current approaches in the literature regarding anchor point placement, highlighting the methodologies used and identifying gaps that this study aims to fill.
Section \ref{sec:materials_methods} outlines the design and functioning principle of the used exosuit, the experimental protocol, the setup, and the analytical and statistical methods employed.
The results presented in Section \ref{sec:results} prove that changing the anchor point placement can significantly impact the range of motion not only of the assisted joint but also on all the lower limbs, with the possibility of both augmentation and reduction. The muscular activation and metabolic data analysis shows that a subject-specific approach should be preferred.

\section{State of the Art}
\label{sec:state_art}

\begin{table*}[ht]
    \centering
    \begin{tabular}{lllll}
         \textbf{Reference} & \textbf{Exoskeleton type}  & \textbf{Body district}  & \textbf{Optimization strategy}  & \textbf{Experimental Validation}\\
         \hline
         Wehner 2013 \cite{wehner2013} &  PAM exosuit & Lower limb & Virtual anchor points & 1 subject\\
         Asbeck 2013 \cite{asbeck2013} & Cable driven exosuit & Lower limb	& Virtual anchor points & 1 subject\\
         Nycz 2015 \cite{nycz2015} & cable driven exosuit & Elbow-hand & Biomimetic placement & No \\
         Guan 2016 \cite{guan2016} & Passive exoskeleton & Lower limb  & Simulation. CF: peak of absolute hip moment & 1 SCI patient \\
         Wei 2018 \cite{wei2018} & Cable driven exosuit & Upper limb & Simulation. CF: man-machine interaction force & 1 subject \\
         Guan 2019 \cite{guan2019} & Passive exoskeleton & Lower limb  & Simulation. CF: peak of absolute hip moment & No \\
         Wu 2019 \cite{Wu2019} & Cable driven exosuit & Elbow & Maximum stiffness & 1 subject \\
         Joshi 2022 \cite{joshi2022} & Passive exosuit  & Elbow  & Simulation. CF: norm of muscle activation & 6 subjects \\
         Bardi 2022 \cite{Bardi2022} & Cable driven exosuit  & Upper limb  & Simulation. CF: power consumption  &  Test bench\\
         Bermejo-Garcia 2023 \cite{BermejoGarcia2023} &  Cable driven exosuit & Lower limb  & Simulation. CF: metabolic consumption  & No \\
         Chen 2024 \cite{chen2024} & Cable driven exosuit  & Lower limb & Genetic algorithm on simulated model  & No \\
         Bonab 2024 \cite{bonab2024} & Cable driven exosuit  & Elbow-wrist & Genetic algorithm on simulated model  & 3 subjects \\
         Prasad 2024 \cite{Prasad2024} & Cable driven exosuit & Lower limb & Simulation. Multiple CFs. & No \\
         Lambranzi 2025 & Clutch-spring exosuit & Lower limb & Simulation. CF: metabolic consumption & 11 subjects \\
    \end{tabular}
    \caption{Anchor points placement strategies in the literature}
    \label{tab:anchor_points_literature}
\end{table*}

Positioning of the actuators in a rigid active exoskeleton comes with a straightforward requirement: the exoskeleton joints should ensure kinematic compatibility, avoiding the misalignment of the exoskeleton’s joints instantaneous center of rotation with respect to the desired position \cite{Sposito2020}.
The two main strategies to guarantee compatibility are mimicking the anatomical joint or using a simplified exoskeleton joint structure to ensure proper alignment between the axes of rotation, such as that of anthropomorphic exoskeletons \cite{naf2019}.
This process is more complex for exosuits: in contrast to rigid exoskeletons, their joint positions are not well defined, and the actuators are distributed along the body \cite{tonazzini2018}.

One of the most essential characteristics of these devices is the way they transfer forces to the human body \cite{xiloyannis2022}, which is crucial to ensure safety and optimize the assistance. Despite its importance, the study of how fabric and elastomers generate load paths is a field that has not been thoroughly explored.

The literature highlights four key aspects to consider when designing force application points in exosuits:
\begin{itemize}
    \item The amount and distribution of applied pressure: Unwanted forces and pressure peaks can cause blisters, bruises, or local blood flow obstruction, which should be avoided, especially in the case of older adults, who have fragile skin that is prone to these phenomena and have a lower tolerance to pain caused by the cuffs' pressure \cite{kermavnar2018}.
    \item The stability of the force application points: Anchor points keep the wearable device in position while actuation forces are transmitted to the body \cite{asbeck2014} and are fundamental to obtaining an efficient and accurate force transmission \cite{asbeck2013}. They must provide a tight and stable physical coupling between the user and the suit, ensuring adaptability to the user's body morphology and comfort \cite{tonazzini2018}.
    \item The routing of the actuators: This aspect is particularly relevant for cable-based exosuits, which need to ensure tension in the cables, maintain safe clearance with the user body, and ensure no interference among the cables \cite{Prasad2024}. One of the first adopted strategies to identify the optimal path for the actuators is that of the non-extension lines, a concept introduced by Iberall \cite{iberall1964} and later applied in the work of Asbeck et al. \cite{asbeck2013,asbeck2014}. These lines are helpful to avoid the actuators' unwanted motion when the subject is moving. Another solution to this problem is the biomimetic approach, so routing the cables in a way that follows the underlying muscles \cite{nycz2015, Li2018, Prasad2024}.
    \item The location of the points where forces are applied: This aspect changes the diagram of assisting forces on the body, influencing muscle response, kinematics, and dynamics.
\end{itemize}

This paper provides a detailed investigation of anchor point effects. All the examples found in the literature are summarized in Table \ref{tab:anchor_points_literature}. To contextualize our approach, we now review existing methods to determine anchor points placement, which can be categorized as follows:
\begin{itemize}
    \item biomimetic placement;
    \item using bony prominences as anchor points;
    \item simulation-based.
\end{itemize}

Biomimetic placement is based on placing the anchor points close to the muscle intersections, allowing the actuators to align with the underlying muscles and follow their paths \cite{nycz2015}.

The virtual anchor points approach was applied to an exosuit powered by Pneumatic Artificial Muscles (PAM) \cite{wehner2013} and later to a cable-driven prototype \cite{asbeck2013,asbeck2014}. In this strategy, the key anchor points are identified at the intersections of bony prominences with the lines of non-extension, which are lines on the skin that do not result stretched during body motion \cite{iberall1964}.

Wu et al. \cite{Wu2019} started from the considerations of Asbeck et al. \cite{asbeck2014} and determined the ideal anchor point as the one with the least deformation of the skin given a force exerted on it.

There are several examples of simulation-based approaches: the majority of them optimize the position of exosuits' anchor points combined with musculoskeletal models \cite{guan2016, guan2019, joshi2022, Bardi2022, BermejoGarcia2023,chen2024,bonab2024}, while others build specific biomechanical models \cite{wei2018, Prasad2024}. These works have different Cost Functions (CF), often based on the simulated energetic expenditure, the comparison of the human-exoskeleton model with the collected data, or the joint moment.

The choice of force application points in an exosuit is ultimately a trade-off between optimality and physical constraints. The modeling approach has the potential to optimize the design process without the need for extensive testing directly on patients and can also provide information on energy mechanisms, allowing the creation of devices that optimize the energy required from their users.
However, a common limitation of all the approaches presented is that they have little to no experimental validation, and often, the simulation approaches compare the simulation results with pre-recorded data to assess the ideal configuration. While this method is convenient, it could lead to conclusions that do not accurately reflect real-world conditions.
This paper suggests that simulations can guide the design of a wearable device and that, in turn, they should incorporate the findings and considerations that emerge from experimental validation and the requirements imposed by space and motion constraints. Starting from these considerations, this study aims to bridge the gap between simulation-based optimization and real-world validation by experimentally analyzing the effects of different anchor point placements on biomechanics and energetics. In this way, we can better understand how anchor point placement affects exosuit performance, providing insights that could improve simulation models and design decisions.

\begin{figure*}[!ht]
\centering

    \begin{minipage}[b]{0.15\textwidth}
        \centering
        \includegraphics[width=\linewidth, height = 6.3cm]{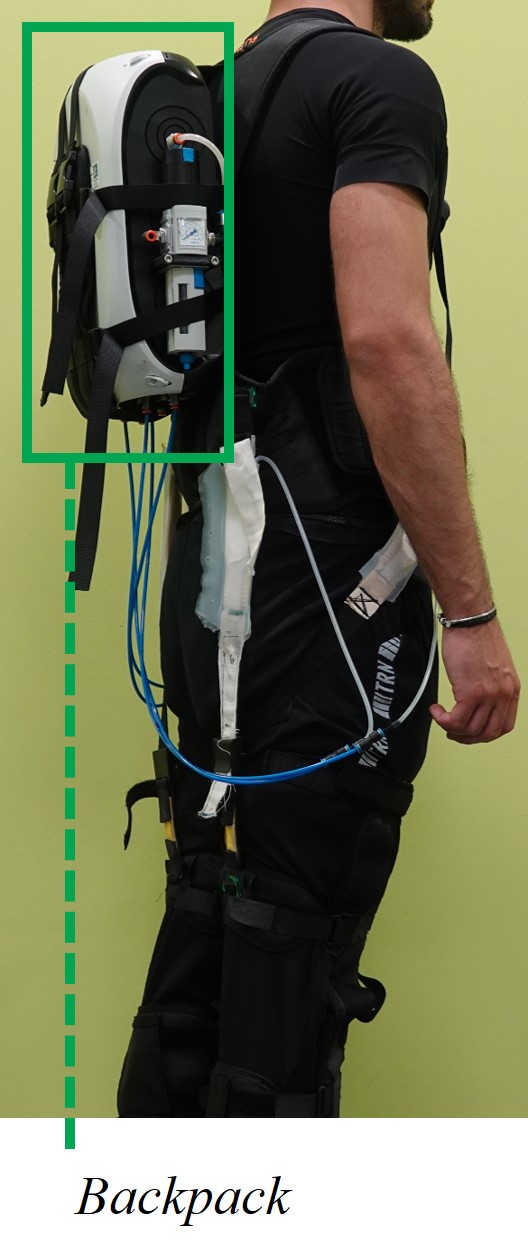}
        \captionof{subfigure}{}
    \end{minipage}
    \hspace{0.01\textwidth} 
    \begin{minipage}[b]{0.15\textwidth}
        \centering
        \includegraphics[width=\linewidth, height = 6.3cm]{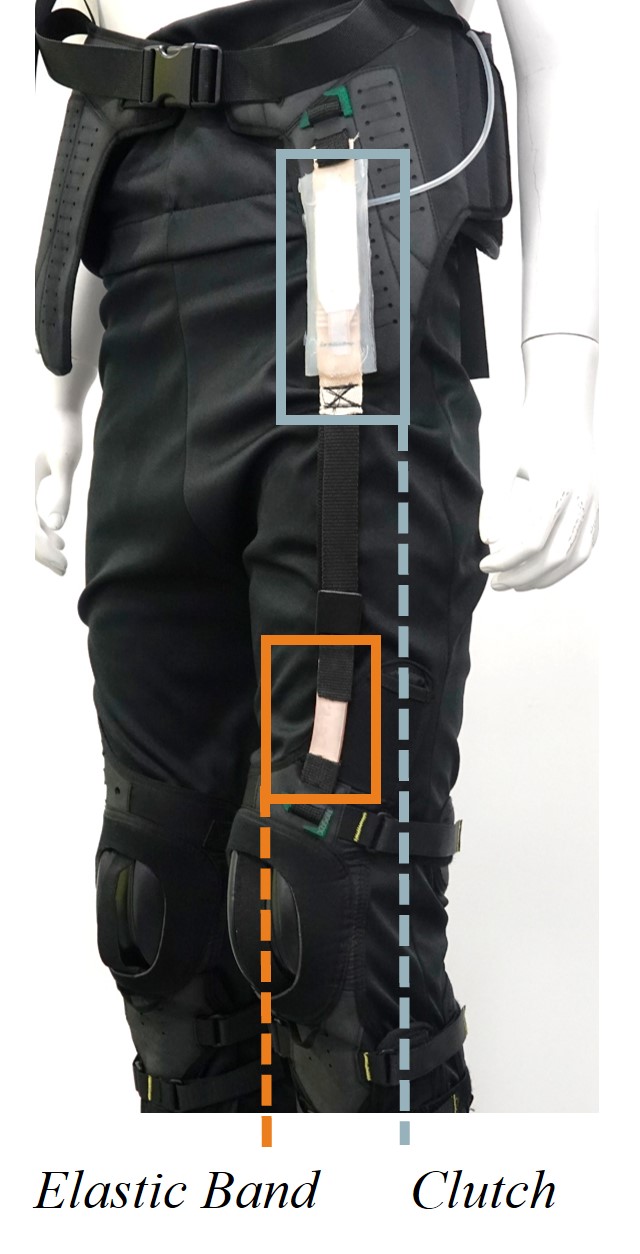}
        \captionof{subfigure}{}
    \end{minipage}
    \hspace{0.005\textwidth} 
    \begin{minipage}[b]{0.65\textwidth}
        \centering
        \includegraphics[width=\linewidth, height = 1.55cm]{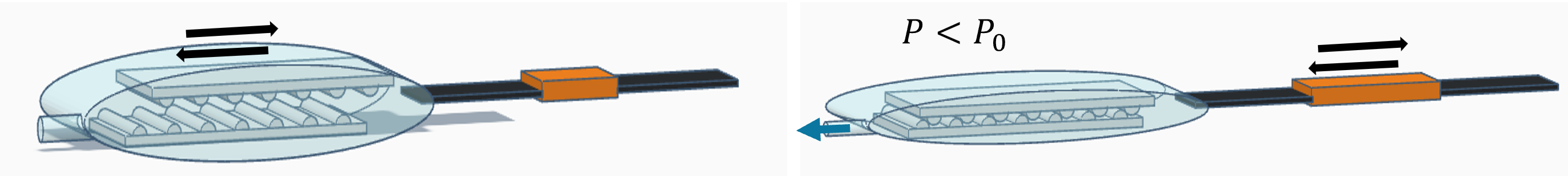}
        \captionof{subfigure}{}
        \vspace{0.2cm} 
        \includegraphics[width=\linewidth, height = 4.4cm]{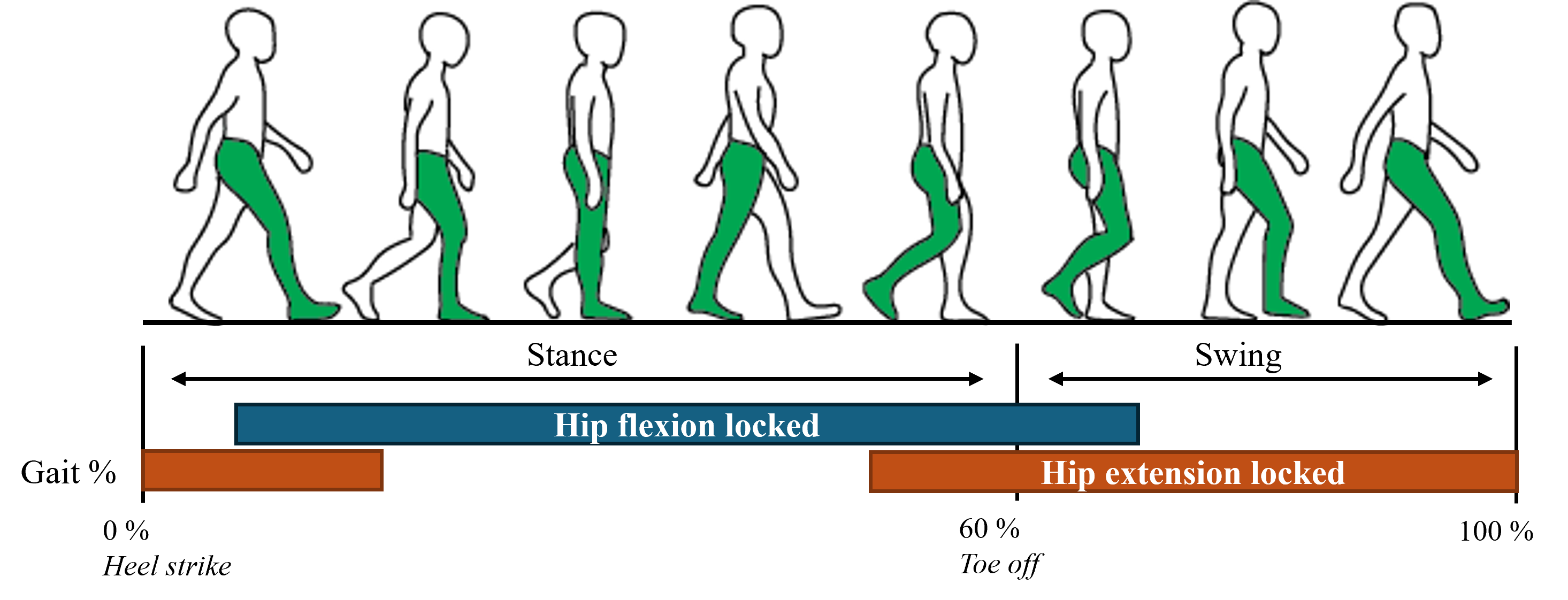}
        \captionof{subfigure}{}
    \end{minipage}

\caption{XoSoft garment and its components. (a) Backpack and posterior view of the system. (b) Clutch-spring series actuator. The belt has several loops that can serve as anchor points. In this experiment, the highest loop, which sits right above the iliac crest, was chosen for all the subjects in the frontal configuration to ensure an initial tension of the actuator. (c) Working principle of the clutch-spring actuator. The clutch comprises two comb-like elements encapsulated in a silicone case. The two combs can slide on each other when the clutch is unlocked, following contractions and extensions. Then, when a pressure $P < P_0$ lower than the atmospheric pressure is created inside the clutch, the two combs lock, and the elastic element will store energy during elongation. (d) Control strategy for the hip flexion and extension actuators depicted during the gait cycle.}
\label{fig:garment}

\end{figure*}

\section{Materials and Methods}
\label{sec:materials_methods}

\subsection{XoSoft exosuit}
\label{subsec:xosoft}

XoSoft Gamma Fig.\ref{fig:garment} is a quasi-passive exoskeleton designed for lower limb assistance, developed during the EU Project XoSoft. 
It is composed of two leggings with straps to secure the garment and attach the actuators, a $4.4 kg$ backpack where the electronics and pneumatic system are stored, two insoles with force-sensitive resistors (FSRs). The system uses an external power supply and an air pressure input.

The actuation strategy relies on storing and releasing elastic energy via selective elongation of an elastic band, which is part of a series actuator comprising a pneumatic clutch and a spring \cite{dinatali2020}.
The clutch contains a spring of negligible stiffness, and is connected in series to a spring element of high stiffness (k = 1.6 N/mm for small deformations). When the clutch is disengaged, it is free to slide and no storage or release of elastic energy occurs. Instead, when we want to store elastic energy, a negative pressure is created inside the clutch, which locks, stretching the elastic element, as depicted in Fig.\ref{fig:garment}(c).
The control system is based on a finite state machine informed by the FSRs in the insoles, which identifies the events that correspond to different phases in a walking cycle.

The system is modular and can assist up to six movements: hip flexion/extension, knee flexion/extension, ankle plantar/dorsiflexion.
In the setup used, it provides symmetric assistance for hip flexion and extension. The actuators are attached to the garment through a dog-bone system, that can be seen in Fig.\ref{fig:garment}(c).
In this experiment, the exosuit assists only the hip, with no support at the knee or ankle. The actuator that helps hip extension has its anchor points fixed, while the attachments for the hip flexion actuator change between the different configurations that are presented in Fig.\ref{fig:configurations}.

Based on previous work on the same device \cite{fanti2022}, the chosen control strategy for storing the elastic energy, as depicted in  Fig.\ref{fig:garment}(d), is:
\begin{itemize}
    \item Hip Flexion: the clutch locks between 5\% - 65\% of the gait cycle, during the stance until the push off;
    \item Hip Extension: the clutch locks between 50\%of the gait cycle - 15\% of the next gait cycle, that corresponds roughly to the swing phase, until the start of extension.
\end{itemize}

\subsection{Experimental Protocol and Setup}
\label{subsec:experimental_protocol}

\begin{figure}[!ht]
\centering
\subfigure[]{\includegraphics[width=0.3\linewidth,height = 5.7cm]{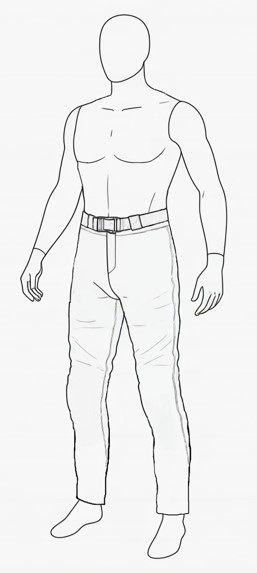}}
\subfigure[]{\includegraphics[width=0.3\linewidth,height = 5.7cm]{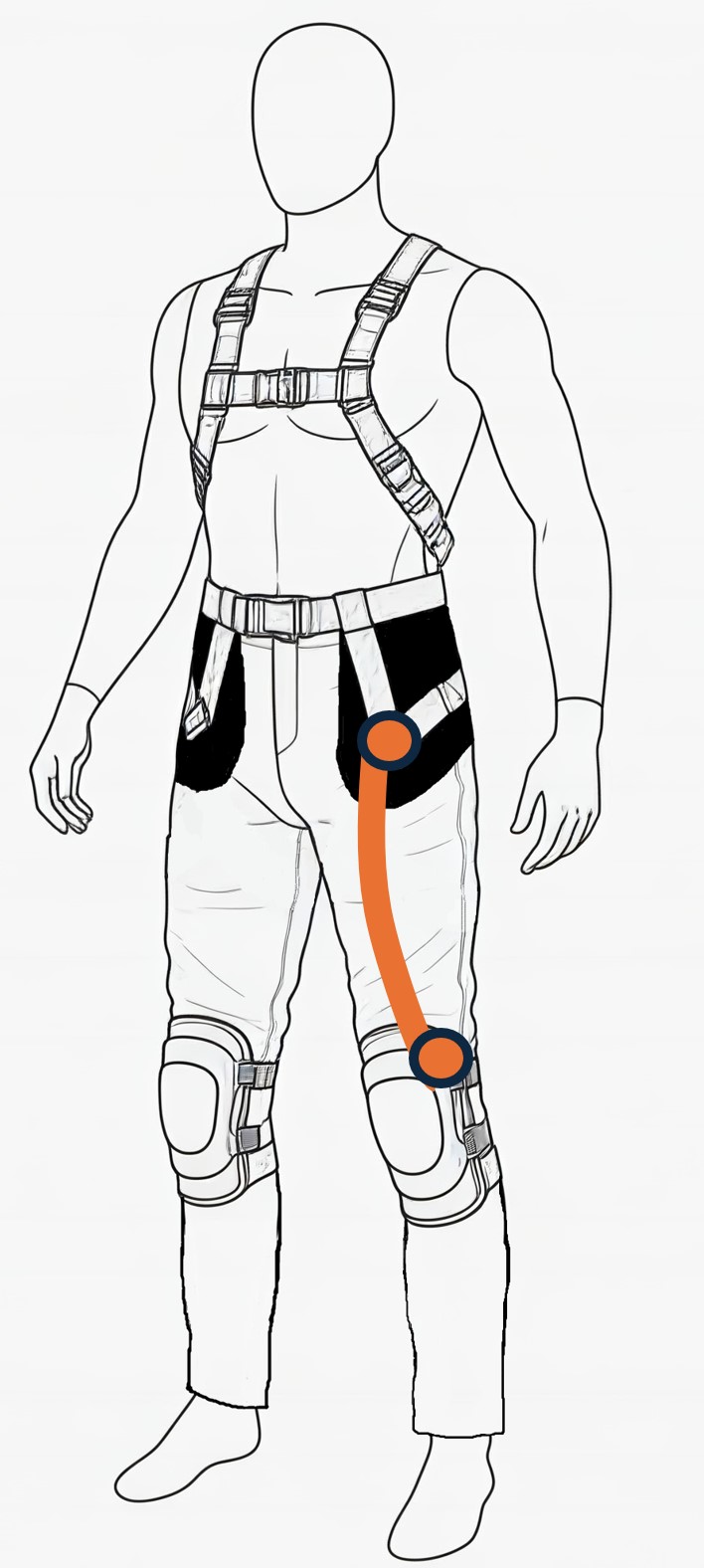}}
\centering
\subfigure[]{\includegraphics[width=0.3\linewidth,height = 5.7cm]{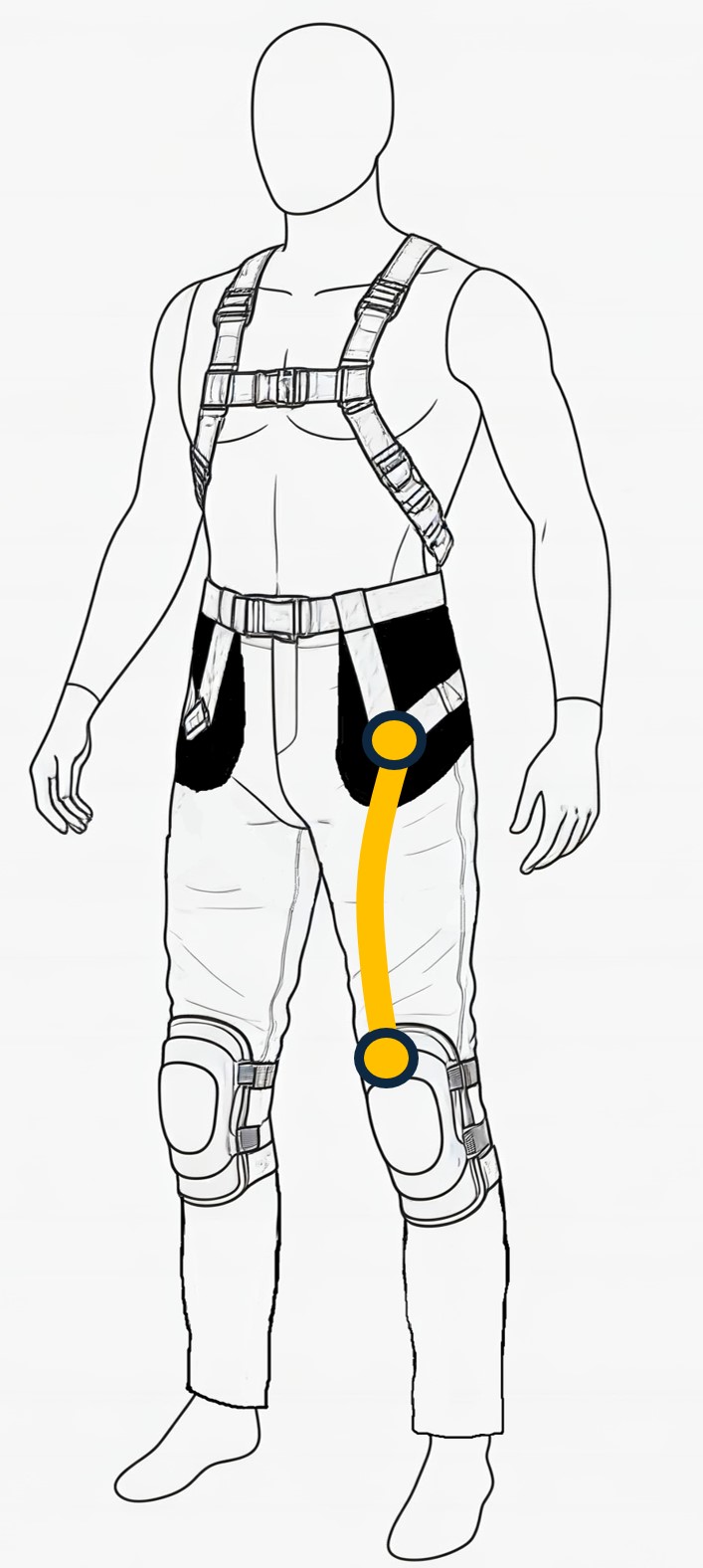}}
\centering
\subfigure[]{\includegraphics[width=0.3\linewidth,height = 5.7cm]{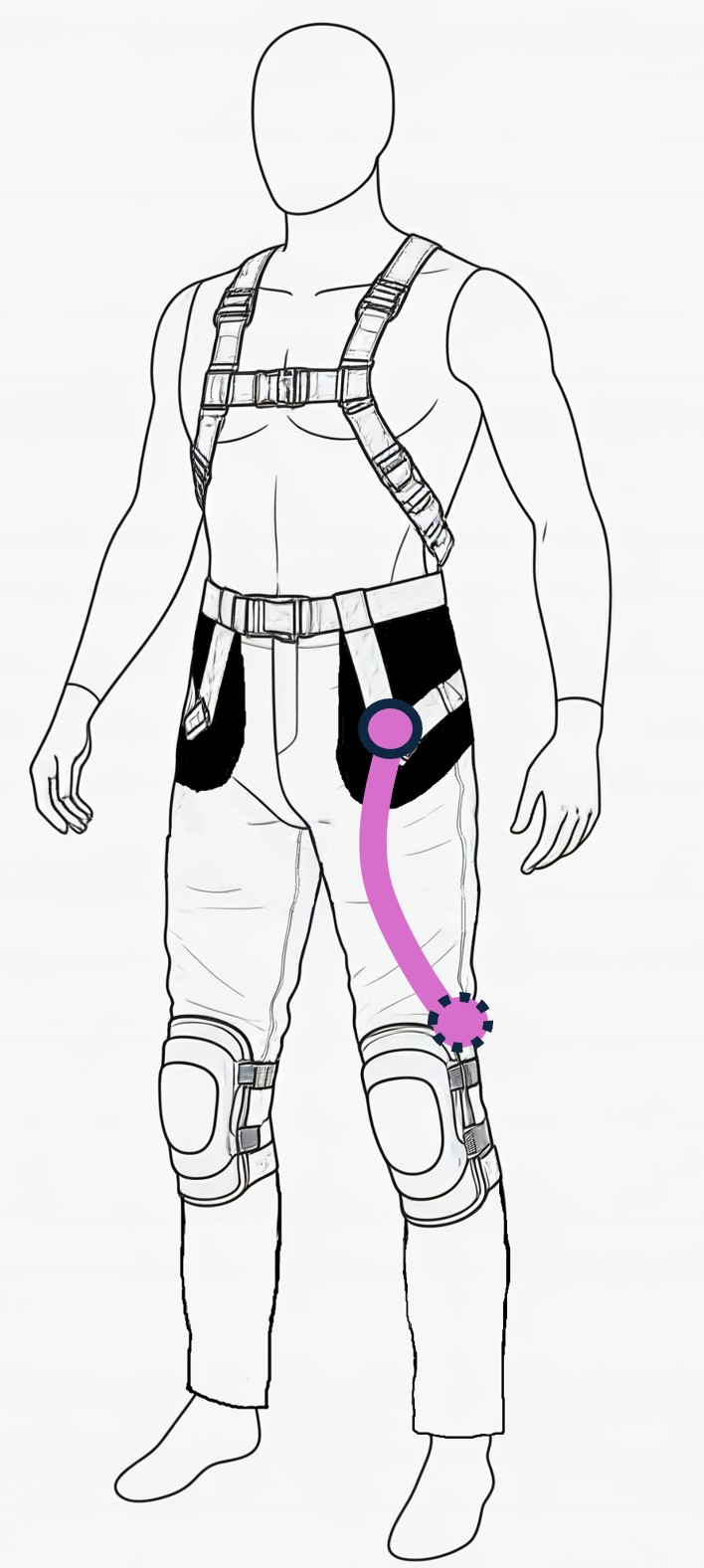}}
\centering
\subfigure[]{\includegraphics[width=0.3\linewidth,height = 5.7cm]{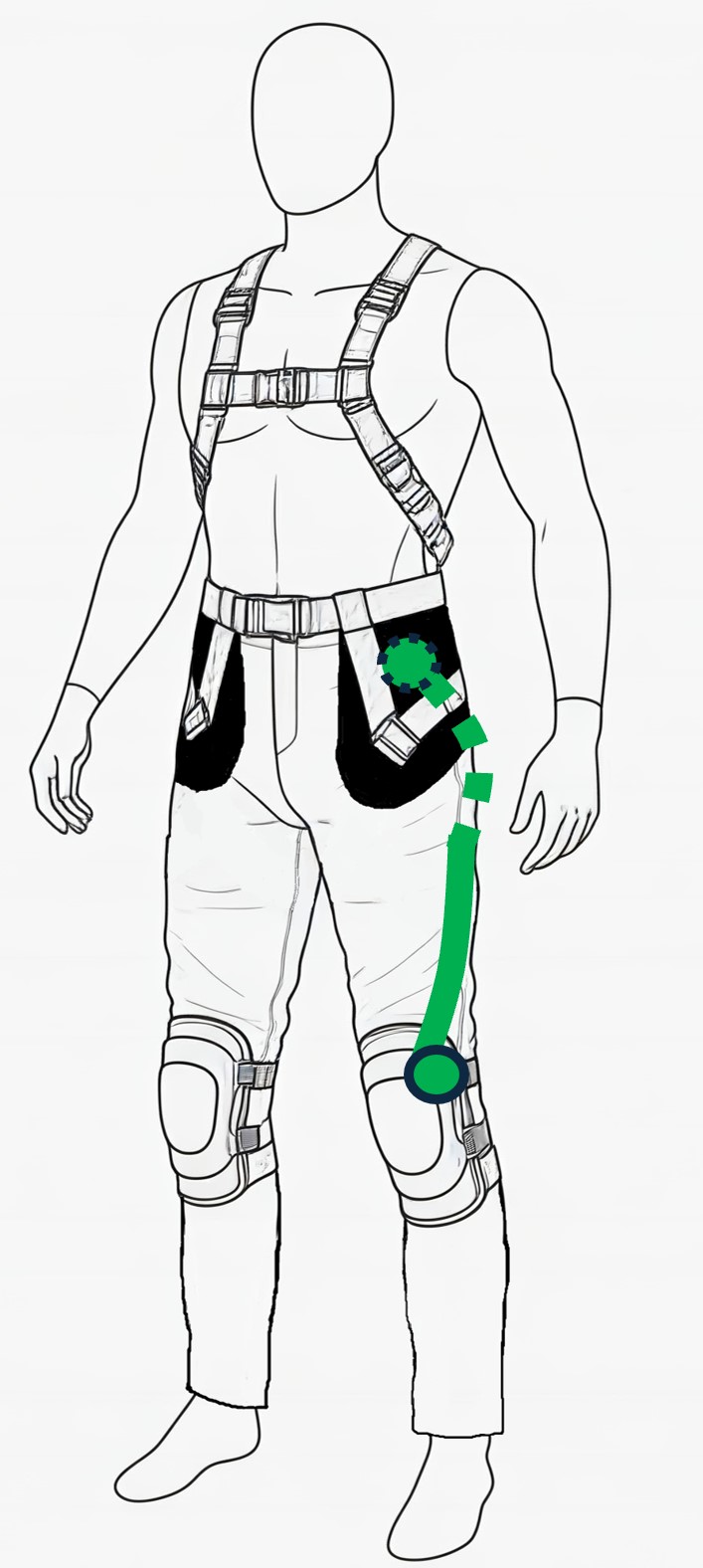}}
\centering
\subfigure[]{\includegraphics[width=0.3\linewidth,height = 5.7cm]{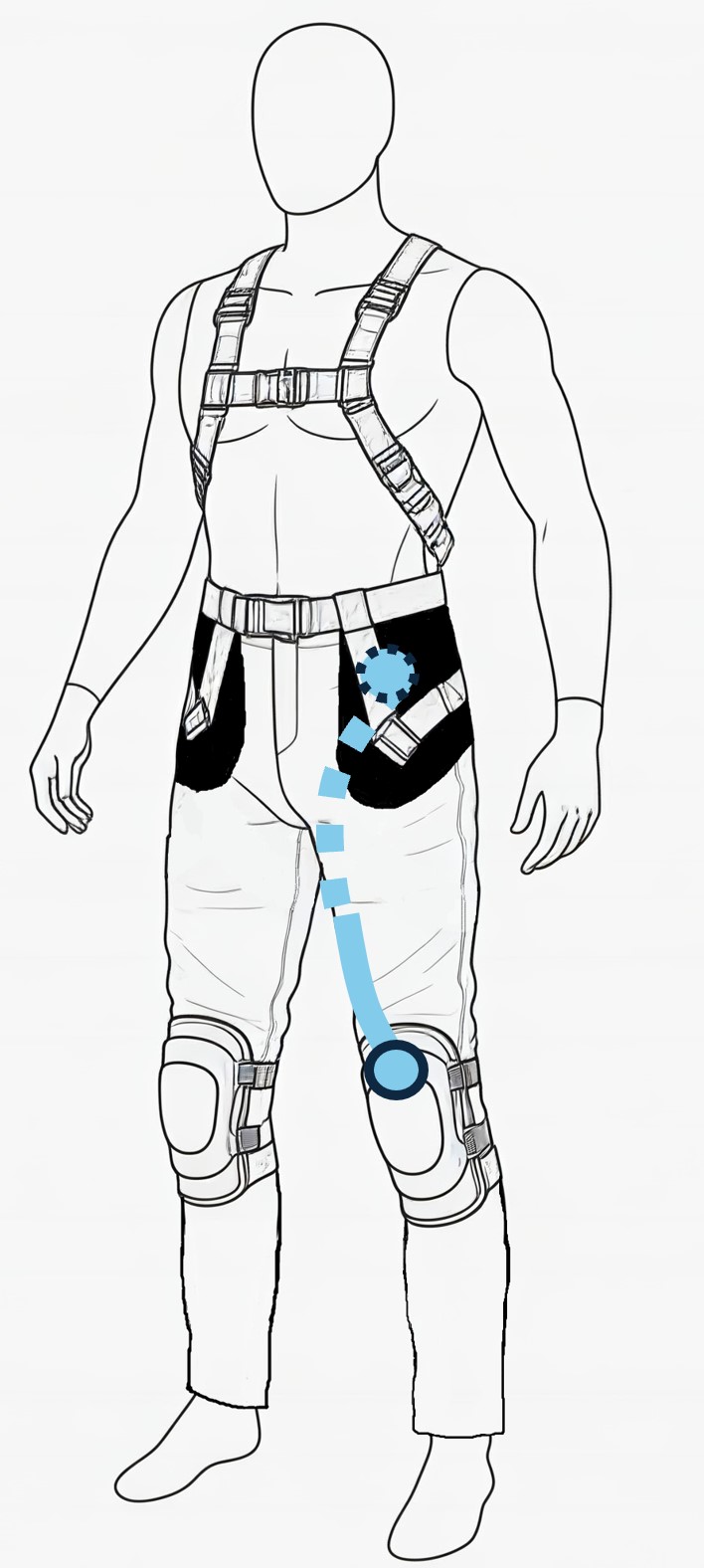}}

\caption{(a) Configuration A: without XoSoft; (b) Configuration B: the anchor points are on the front of the hip and on the front-side of the knee; (c) Configuration C: the anchor points are on the front of the hip and on the center of the knee front; (d) Configuration D: the anchor points are on the front of the hip and on the back of the knee; (e) Configuration E: the anchor points are on the back of the hip and on the knee front-side; (f) Configuration F: the anchor points are on the back of the hip and to the front-center of the knee, following the path of the iliopsoas.}
\label{fig:configurations}
 
\end{figure}

The study involved 11 participants, 5 females and 6 males. The participants were aged between 22 and 33 years (M = 28.45,SD = 2.18), had a height varying from 164 cm to 192 cm (M = 175.63,SD = 8.38), and a weight between 55 kg and 82 kg (M = 66.54,SD = 8.19).
Inclusion criteria required the subjects to be healthy individuals without walking difficulties, cardiac or respiratory conditions, as well as tendon, muscle and bone injuries.
This study was performed in accordance with the protocol approved by the Ethics Committee of Liguria, Italy (protocol number: 001/2019). 

The number of subjects was limited by the long duration of each session and the large number of conditions tested, which made larger recruitment impractical within available resources. Despite the small sample of healthy adults, this remains one of the largest trials of its kind to date. Young participants were selected to assess the actuation strategy under controlled conditions, minimizing confounding factors like muscle weakness or pathological gait. This allowed for a clearer evaluation of mechanical, muscular, and energetic effects. While larger clinical trials are planned, the primary goal here was to validate anchor point concepts and actuation effects. Notably, the same XoSoft platform has previously been tested on patients \cite{dinatali2020}, though recruitment remains challenging due to the difficulty of identifying homogeneous clinical populations.

Experimental acquisitions begin with a two-minute adjustment period, during which participants got familiar with wearing the exoskeleton and the sensors.
Before the trials, participants undergo a 6-minutes resting metabolic consumption measurement while sitting, which was used as a baseline value.
The experiment consisted of walking on a treadmill at a constant speed of 5km/h for 8 minutes, for a total of six trials: the first without the use of the exoskeleton (Configuration A), then using XoSoft in five different configurations (Configuration B-F), that were randomized between users. These configurations are characterized by different placements of the actuators: they were the possible combination of two anchor points at the hip level, anterior and posterior, and three at the knee level, posterior, lateral anterior and central anterior, as shown in Fig.\ref{fig:configurations}.
A rest phase of at least 5 minutes was ensured between each trial to avoid fatigue.
To ensure consistent reattachment of anchor points across trials, alignment was guided by anatomical landmarks. The patella was the anatomical reference for aligning the knee anchor point, enabling consistent realignment of the pants, which were prone to minor downward displacement between trials.
The backpack, which contains the hip anchor points, was aligned using the iliac crest as a reference and repositioned consistently between trials without altering the length of its adjustable components.

The subjects were instrumented with:
\begin{itemize}
    \item a motion capture system (MTw Awinda 3D Wireless Motion Tracker, Movella Inc,  Henderson, NV, USA) that captured the kinematics of the lower body at a sampling rate of 100 Hz;
    \item  an 8-channel Wi-Fi transmission surface electromyography (FreeEMG 300 System, BTS, Milan, Italy) was used to acquire the surface myoelectric signals (sEMG) at a sampling rate of 1,000 Hz. It recorded signals from rectus femoris, tensor fasciae latae, biceps femoris and erector spinae. The sensors were placed on both sides of the body, according to the SENIAM guidelines;
    \item an ergospirometer (K5, Cosmed Srl, Rome, Italy) to measure oxygen consumption and carbon dioxide production. The sensor was used in the breath by breath mode.
\end{itemize}

\subsection{Data analysis and statistical validation}
\label{subsec:analysis_stat}
Data were processed using MATLAB software (MATLAB 2024b, MathWorks, Natick, MA, USA).

\subsubsection{Kinematic data}
Since the experiments were performed on a treadmill at a preset speed and with limited walking space, the classic space-time parameters are subject to constraints that make them different from free walking, so they will not be discussed in detail. 
Instead, the kinematic analysis focuses on the angles of the joints of the lower limb along the sagittal plane. The preprocessing consists of segmentation of the joint angles vectors, derived from the IMU data, that was shifted by a fixed number of samples (15 samples) to compensate for the delay of the acquisition system.

To evaluate the similarity between the joint angles with and without the exosuit, we used Dynamic Time Warping (DTW), introduced by Sakoe and Chiba \cite{sakoe1978}. DTW computes the optimal alignment between the sequences by minimizing the cumulative distance between corresponding points.
It is defined recursively as:
\[
\text{DTW}(i, j) = \|x_i - y_j\| + \min 
\begin{cases}
\text{DTW}(i-1, j), \\
\text{DTW}(i, j-1), \\
\text{DTW}(i-1, j-1),
\end{cases}
\]

where: \(\|x_i - y_j\|\) is the distance between the \(i\)-th element of \(X\) and the \(j\)-th element of \(Y\), \(\text{DTW}(i, j)\) is the cumulative DTW distance up to the \(i\)-th element of \(X\) and the \(j\)-th element of \(Y\). The indices i and j range from $1$ to $n$ and $1$ to $m$, respectively, where $n$ is the length of sequence $X$ and $m$ is the length of sequence $Y$.
The final DTW distance between the two sequences is given by $ d = \text{DTW}(n, m).$

\subsubsection{EMG data}
EMG data were preprocessed according to the estimated metric, that respectively related to mean frequency and muscle activation.

Mean frequency (MNF), as defined by Phinyomark et al. \cite{Phinyomark2012}, is the average frequency of an EMG power spectrum. It is calculated as the sum of the product of the EMG power spectrum $P_j$ and the frequency $f_j$ of the frequency bin $j$ divided by the total sum of the power spectrum. $M$ is the next power of 2 from the length of EMG data in time-domain.

\begin{equation}
    \text{MNF} = \frac{\sum_{j=1}^Mf_jP_j}{\sum_{j=1}^MPj}
    \label{eq:MNF}
\end{equation}

A decrease over time in the spectral mean frequency is associated with muscular fatigue. It is often adopted in short, high-intensity exercises but has some examples of use also in cyclical dynamic contractions, accounting for the duration of the experiment by considering the slope of the regression line that fits maximum values of MNF in a time window \cite{Thongpanja2013}.

To compute MNF, the EMG data were preprocessed with a second order Butterworth bandpass filter 20-300Hz, then the MNF was computed Eq. \ref{eq:MNF} in non-overlapping windows of 1000 samples. Then, the vector containing all the MNFs was linearly interpolated to obtain the slope of the first-degree fitting curve. A negative slope represents a negative shift in the power spectrum over time.\\

To analyze muscle activation, EMG signals were first bandpass filtered using a second-order Butterworth filter (20–300 Hz), then rectified, low-pass filtered with a second-order Butterworth filter at 5 Hz, and finally normalized to the 95th percentile of the signal amplitude.
The data were then segmented in gait cycles using the contact of the foot with the ground, obtained from the IMU data, as a reference.

The metrics assessed are the percentage variation $PV$ between Configuration A, without XoSoft, and the others, the integral-based assistance index $I_{trend}$ introduced by Fanti \cite{fanti2024}, and the co-contraction index $CCI$ introduced by Falconer and Winter \cite{Falconer1985}.

The integral-based assistance index $I_{trend}$ allows the evaluation of the exoskeleton’s impact on the user musculature during the entire task duration. It needs a precise synchronization of the compared signals, performed by taking the contact of the foot with the ground as a reference and resampling the two signals.
$I_{trend}$ can be defined as follows:
\begin{equation}
    I_{trend}(i) = \frac{EMG_{NoExo}(i)*\delta t - EMG_{Exo}(i)*\delta t}{\sum_{i=1}^{N}EMG_{NoExo}(i)*\Delta t} *100
    \label{eq:itrend}
\end{equation}
where $\delta t$ represents the sampling time, $\Delta t$ is the duration of the task and $N$ is total number of samples $i$.
The difference between the EMG activities in each sample is normalized by  $EMG_{NoExo}$ integral, which allows the contribution of the exoskeleton to be evaluated at each stage of the task. Positive $I_{trend}$ values reflect a net reduction in muscular activation provided by the exosuit, while negative values may suggest increased effort or suboptimal force transmission. In this case, this metric allows us to have a clearer vision of how the different configurations influence muscular energy expenditure.\\

Co-Contraction Index (CCI) was computed according to Eq. \ref{eq_cci} for the rectus femoris and the biceps femoris; these are the only coupled antagonistic muscles considered since they are respectively a powerful hip flexor and extensor, both accessible to surface EMG sensors.
\begin{equation}
    \text{CCI(i)} = 2\frac{min(abs(EMG_{m}(i)),abs(EMG_n(i)))}{EMG_m(i)+EMG_n(i)}
    \label{eq_cci}
\end{equation}
$EMG_m$ and $EMG_n$ represent EMG signals from an antagonist muscle pair, where both signals were resampled. For each instant, the numerator of the fraction is the  EMG signal with the lowest absolute magnitude in that sample.
Muscle co-contraction causes an increase in joint stiffness, which in turn improves the stability and accuracy of the movement.
At the single-joint level, muscle co-contraction is seen in healthy subjects in different actions, from steady-state tasks to quick movement and force production tasks. 
Populations with impaired motor abilities commonly show increased levels of muscle coactivation \cite{Latash2018}.\\

\subsubsection{Metabolic consumption}
The estimation of the metabolic cost is based on the measurement of the oxygen consumption $VO_2$ and the carbon dioxide emission $VCO_2$,  according to Eq. \ref{eq:NMR} used to derive the normalized Net Metabolic Rate (NMR) expressed in watts [W], adapted from the work of Brockway \cite{Brockway1987}.

\begin{equation}
\text{NMR} = \frac{16.58\;VO_2+4.52\;VCO_2 - MR_{rest}}{W_{sub}+W_{exo}}
\label{eq:NMR}
\end{equation}

NMR is obtained by subtracting the rest metabolic rate $MR_{rest}$ from the metabolic rate of the experiment and normalizing it for the weight of the subject $W_{sub}$ and, where applicable, by the weight of the XoSoft backpack $W_{exo}$.

$MR_{rest} = 16.58\;VO_{2,rest}+4.52\;VCO_{2,rest}$ is the metabolic rate of the 6 minutes of rest that were computed before the experiment and is used to create a baseline measurement of the basal metabolic consumption of each subject.
NMR is computed considering only the last two minutes of activity, when the energetic consumption reaches its plateau.\\

\subsubsection{Statistical analysis}
The approach adopted for an initial analysis used the Friedman test to form a comparative assessment of the significance of the results for different configurations. Subsequently, the Wilcoxon signed rank test Correction was performed to compare the configurations pairwise. These tests were selected since: 
\begin{itemize}
    \item we could not assume independence between the different configurations,
    \item we have a small population,
    \item we wanted to compare different conditions applied to the same subjects.
\end{itemize}
The p-values were not corrected for the multiple comparison problem due to the study's exploratory nature and the limited number of participants.
In upcoming investigations, we will reduce the number of specific comparisons with a larger sample size to enable the adoption of p-value corrections.

In the cases of data incompleteness, the Friedman test was used only on the subset of subjects with data for all conditions; then the pairwise Wilcoxon test was used considering overlapping subjects for each pair. These cases are explicitly reported in the results section.

To assess whether performance was influenced by trial order, for example due to insurgence of fatigue, the data were reorganized according to the sequence position of each trial and a Friedman test was performed. A significant result would indicate that performance varied systematically across trial positions, suggesting an effect of trial order. A non-significant result would suggest no such effect.

\section{Results}
\label{sec:results}

\begin{figure*}[!ht]
\centering
\subfigure[]{\includegraphics[clip, trim={25 30 0 5}, width=0.32\linewidth, height=9cm]{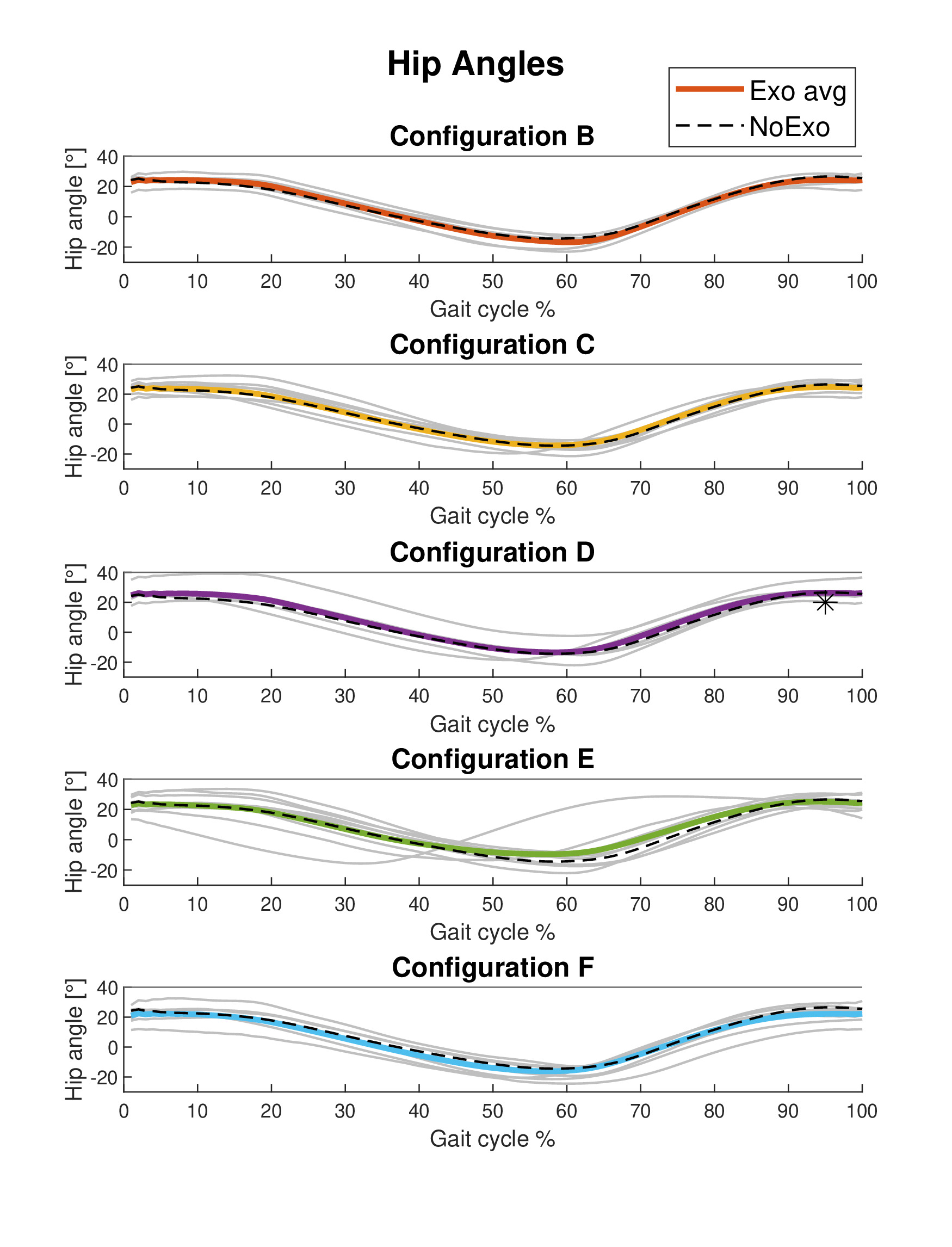}}
\centering
\subfigure[]{\includegraphics[clip, trim={25 30 0 5},width=0.32\linewidth, height=9cm]{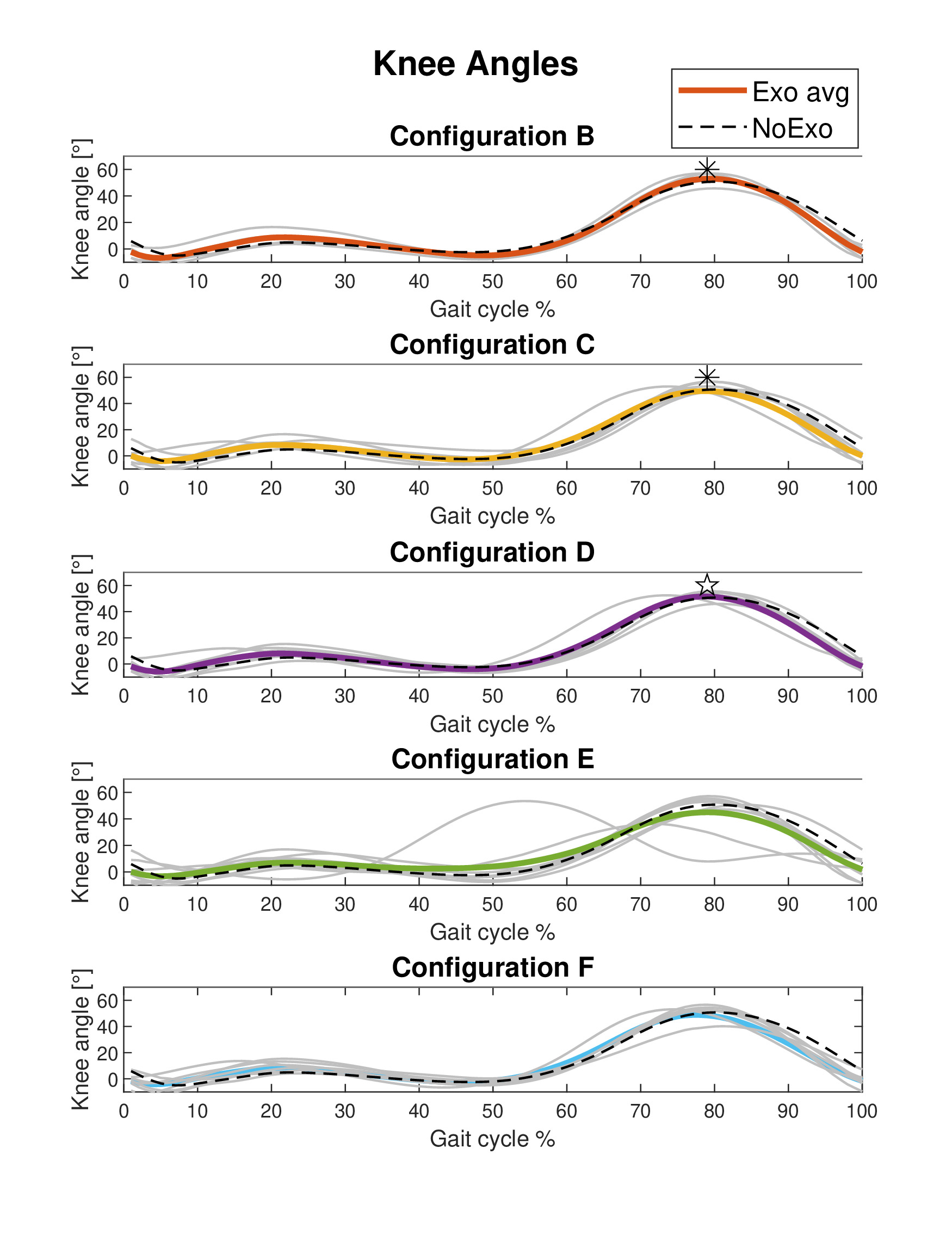}}
\centering
\subfigure[]{\includegraphics[clip, trim={25 30 0 5},width=0.32\linewidth, height=9cm]{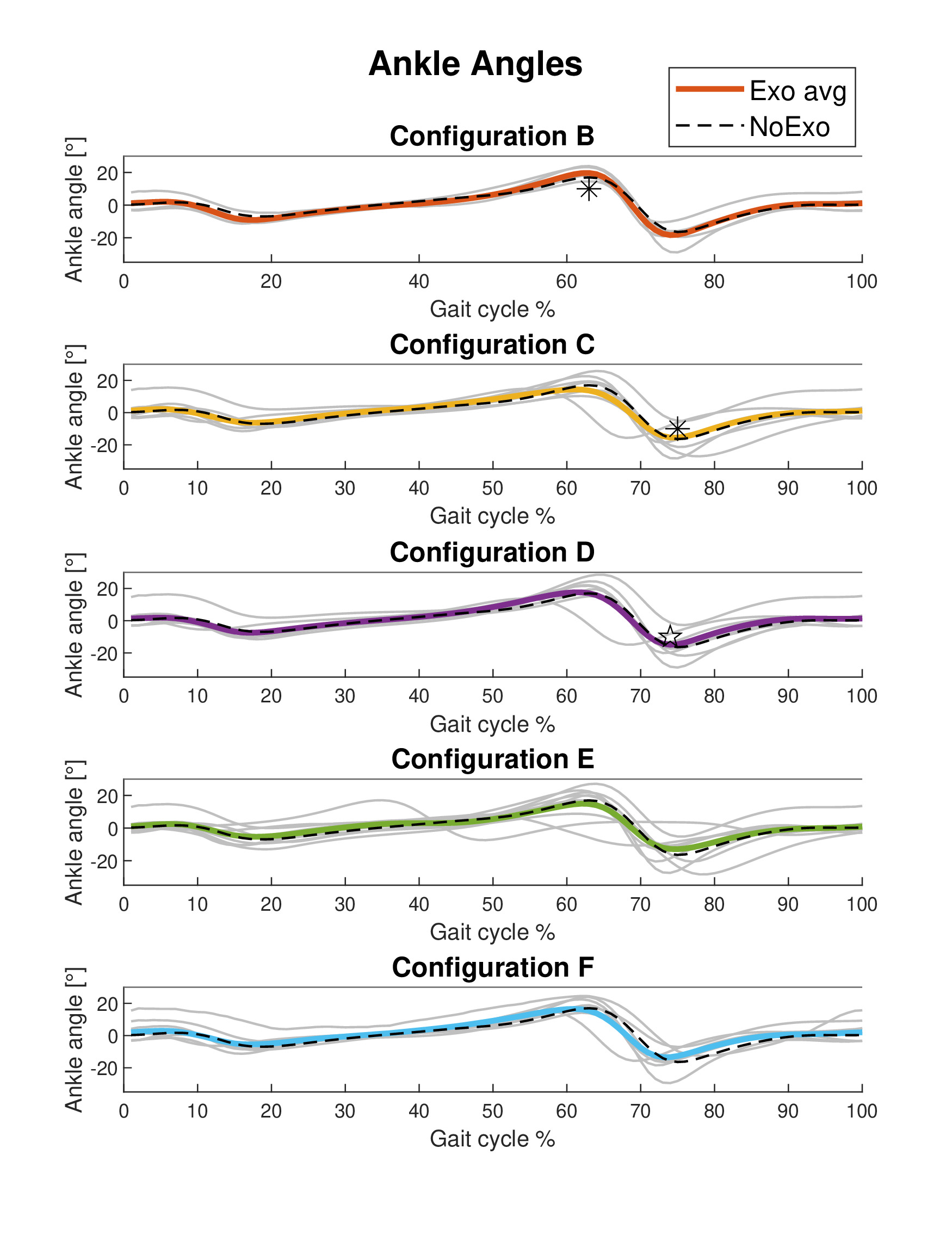}}

\caption{Average joint angles and standard deviations of (a) hip, (b) knee and (c) ankle, for the different configurations compared to Configuration A. The significance reported $*  p < 0.07$, $\smallstar  p < 0.05$,  is in correspondence of the peaks where there is a significant difference between the considered configuration and Configuration A.}
\label{fig:kinematic_analysis}
 
\end{figure*}

\subsection{Kinematic Analysis}

\begin{figure*}[h]
\centering
\subfigure[]{\includegraphics[clip, trim={15 15 15 5}, width=0.32\linewidth, height=6.5cm]{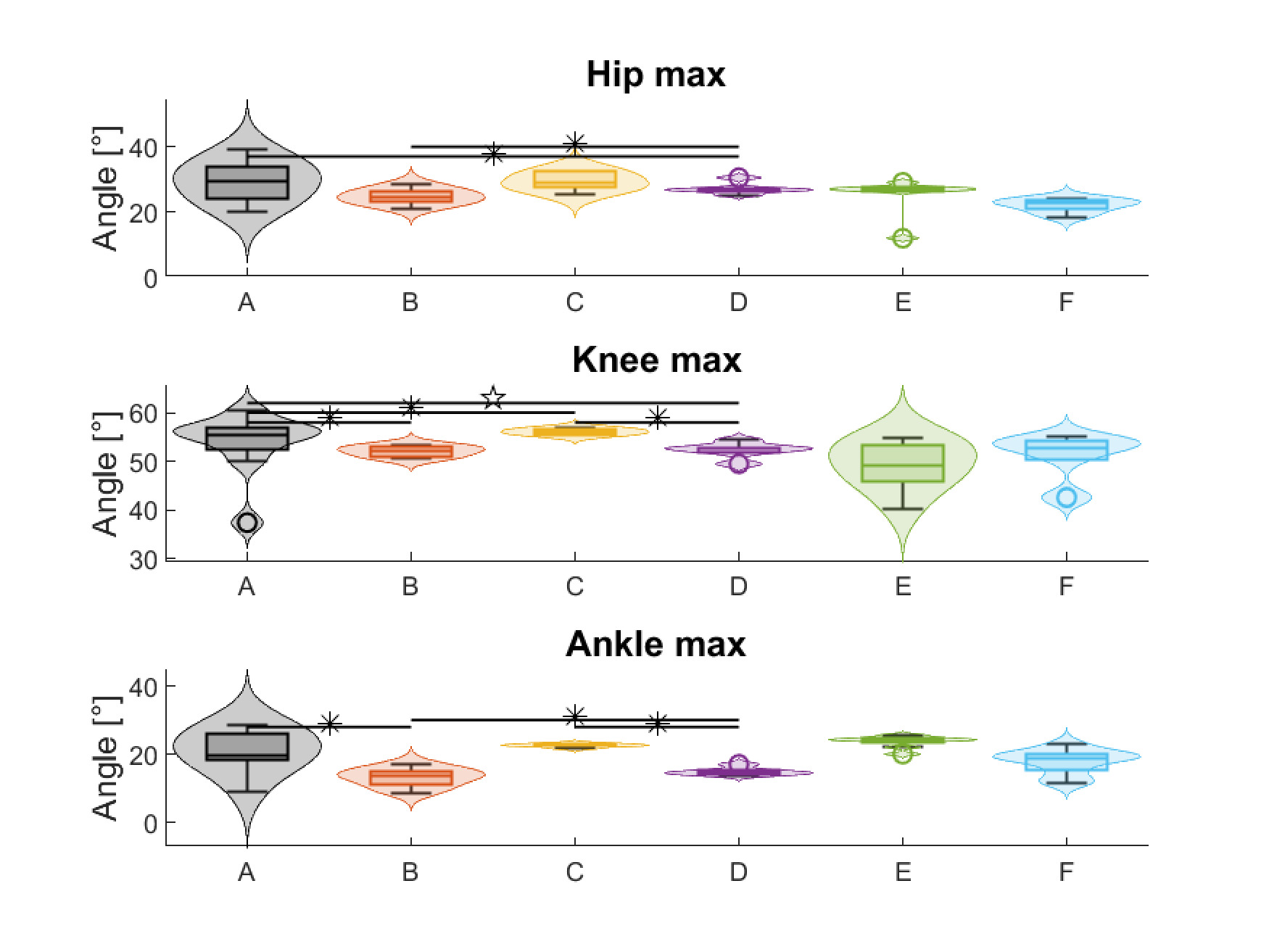}}
\centering
\subfigure[]{\includegraphics[clip, trim={15 15 15 5},width=0.32\linewidth, height=6.5cm]{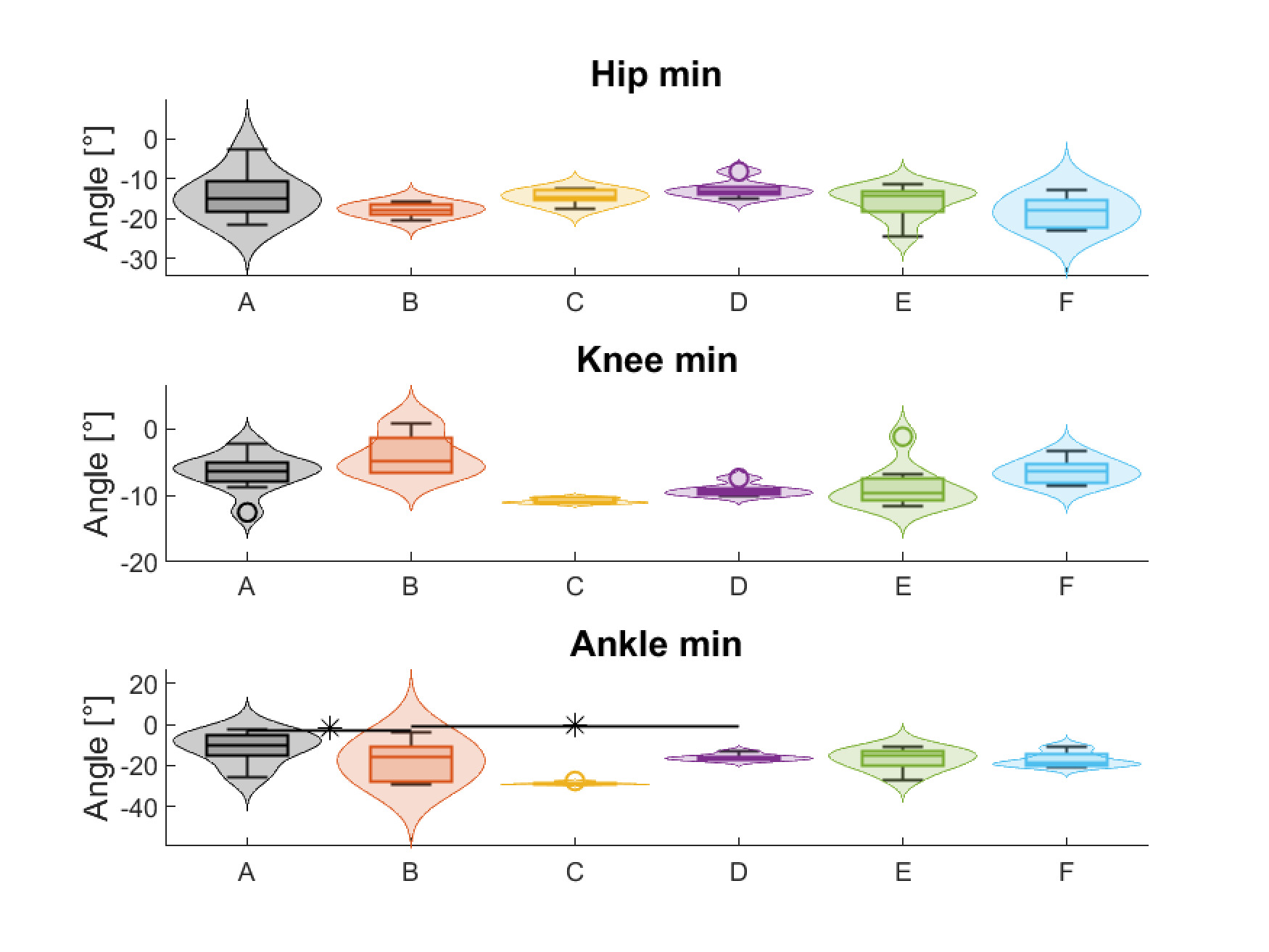}}
\centering
\subfigure[]{\includegraphics[clip, trim={15 15 15 5},width=0.32\linewidth, height=6.5cm]{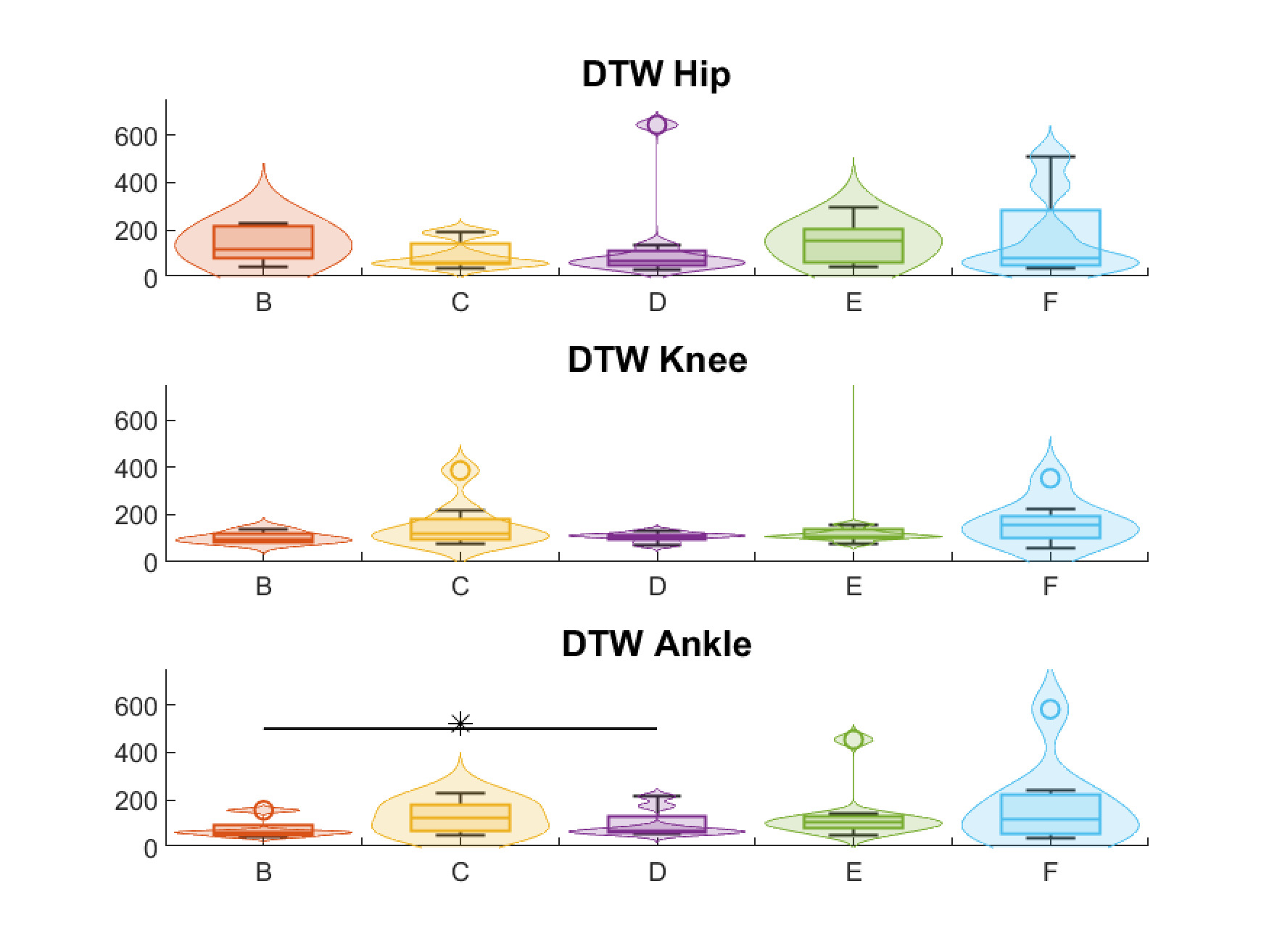}}

\caption{Distributions of the joint curve features for all the subjects, respectively of the (a) maximum peak value (b) minimum peak value, (c) DTW value. Violin plots illustrate the distribution shape but may introduce visual artifacts due to interpolation; boxplots are provided for accurate reference of central tendency and variability. The significance reported $*  p < 0.07$, $\smallstar  p < 0.05$ refers to the pairwise Wilcoxon test. (a) An increase in the maximum peak values represents an increase in the hip flexion, knee flexion and ankle dorsiflexion range of motion. (b) A decrease in the minimum peak values represents an increase in the hip extension, knee extension and ankle plantarflexion range of motion. (c) Higher DTW values indicate greater deviation from baseline joint kinematics, which may reflect less natural or altered movement patterns.}
\label{fig:kinematic_analysis_2}
 
\end{figure*}

The results of the kinematics analysis of the joint angles for the lower limb joints are presented in Fig.\ref{fig:kinematic_analysis}.
The figure shows the average joint angles and the average angles for each subject in the different configurations. As Fig.\ref{fig:kinematic_analysis} shows, even though only the hip flexion and extension were actuated, the introduction of the exosuit had an impact on the knee and ankle kinematics as well, indicating a possible adaptation of the gait to the external force introduced by the elastic element.
The following feature analysis compares values extracted from the curves, specifically the maxima and minima for each subject. Dynamic Time Warping (DTW) is also used to compare each configuration with its corresponding one without the exoskeleton. Statistical analysis accounts for inter-subject variability through pairwise Wilcoxon tests and subject-by-subject correlation studies to assess the temporal order of the data. 
The effects of the introduction of the exoskeleton on the curve peaks are reported in Fig.\ref{fig:kinematic_analysis_2}. The percentage comparisons refer to the median values computed over all the subjects. Configuration A exhibits greater variability in peak values, likely due to being the only configuration with complete data. In contrast, the broader distributions observed in the other configurations may reflect increased variability introduced by interaction with the actuators. On average, the hip flexion angles are smaller for all the configurations where XoSoft is introduced, with the biggest reduction for Configuration F (22.64\textdegree) of 23.56\% with respect to the baseline (29.62\textdegree).
For hip extension, the minimum peak, configurations B and F present an increased range of motion, while the others a reduced one; the two configurations D and F, are those that present the biggest difference: Configuration D (-13.01\textdegree) has a reduction of the 12.4\% of the extension, while Configuration F (-17.8\textdegree). has an increase of the 19.8\%.
For the knee, Configuration C is the only one to present an increase in the range of motion of both the flexion and extension. Configuration E has the biggest reduction of the knee flexion (49.25\textdegree) by 11.26\%,  with respect to the baseline (55.5 \textdegree), while for the extension Configuration C (-10.98), that has a bigger extension by 81.7\%.
For ankle plantar and dorsiflexion both Configuration C and E see an increase in the overall range of motion; for dorsiflexion, Configuration E (24.16\textdegree), has an increase of 22.3\%, while Configuration C has an increase in the plantarflexion range of motion of 177.4\%.
It is interesting that the different configurations were designed to assist hip flexion but resulted in bigger differences in the ranges of motion of the other joints.

Overall, the changes in angular displacements  underline the fact that, due to the exchange of forces with the exosuit, the walking pattern of the user has been slightly modified. With the few notable exceptions already described, most configurations present variations within 2°. This is the threshold defined by Winter for inter-subject variations \cite{winter1991}, and these changes can be considered as compensatory effects for the elastic forces, so the changes in ranges of motion are almost negligible compared to natural variations, in accordance to what has been reported in previous studies \cite{DiNatali2023}.

The Wilcoxon sign-rank test was performed on the set of maximum and minimum peak values for all the subjects. The results are reported in Fig.\ref{fig:kinematic_analysis} and Fig.\ref{fig:kinematic_analysis_2}(a),(b), comparing each Configuration to the one without wearing XoSoft. A higher threshold for significance was chosen to report results that, with a higher population, could turn into actually significant results.
For Configuration B, the peaks differed from the baseline for the knee flexion (p = 0.0625) and the ankle peak dorsiflexion (p = 0.0625).
Configuration C had similar results for the knee flexion (p = 0.0625) and also had a significant difference for the ankle plantar flexion (p = 0.0625).
Configuration D had the greatest difference with respect to the baseline, with significance for the hip flexion (p = 0.0625), knee flexion (p = 0.03125), and ankle plantar flexion (p = 0.03125).
Configurations E and F did not show statistically significant differences even though their average curves have the most different peaks than the reference curve.
To assess the impact of configuration order on joint behavior, a subject-by-subject correlation analysis was performed using the joint maxima and minima. Subject 5 exhibited very strong correlations between knee extension, with $\rho = 1.0, p = 0.01$, and knee flexion $\rho = -0.9, p = 0.08$ for knee flexion. This indicates a consistent trend in this subject, where extension increases, and flexion decreases across configurations. Subject 12 showed a similar, though less pronounced, pattern in knee flexion ($\rho = -1.0, p = 0.08$), suggesting the presence of a comparable trend. However, these high correlations were observed only in a limited number of subjects and only for the knee joint, therefore suggesting a limited impact of the order of execution on the joint kinematics.

The Dynamic Time Warping of the average joint angles was computed to compare the similarity of these curves to Configuration A. The DTW computation was performed by comparing each curve to the corresponding reference curve from the same subject. The results are presented in Fig.~\ref{fig:kinematic_analysis_2}(c). For the hip, the Configurations that result from their median being more distant from the one without exosuit are B (118.3) and E (157). This changes for the knee and ankle, which both present a bigger median distance of the curves in Configurations C (121.5, 122.1) and F (157.2,119.7). With the exception of an outlier, Configuration D is the one consistently closer to Configuration A.
A pairwise Wilcoxon test shows a statistical difference in the results on the ankle DTW between Configuration B and D. A correlation analysis was used to assess the impact of the experimental order subject by subject, that showed a positive correlation ($\rho = 0.9, p = 0.08$) only for Subject 10.

These results are consistent with the statistical test results, suggesting that the significant difference in the peaks could be related to a higher distance between the two curves. The Euclidean distances also reflect the biggest differences presented by the knee and ankle curves with respect to the hip ones.

\subsection{Muscular Activity Analysis}

\begin{figure}[!h]
    \centering
    \includegraphics[clip, trim={15 10 15 10},width=0.95\linewidth]{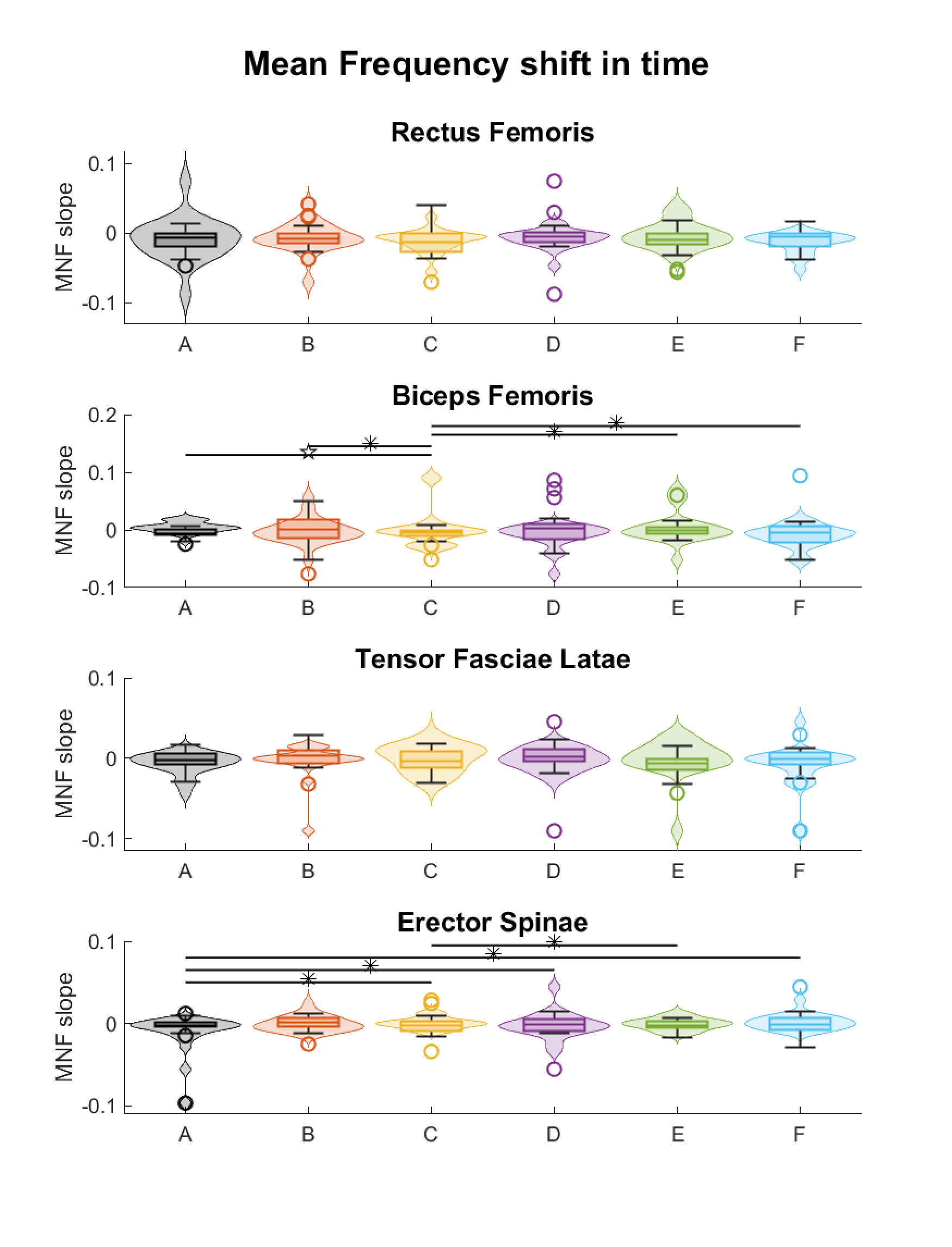}
    \caption{Mean frequency results: the violin plots represent the mean frequency shift over time for the four muscles analyzed in the experiment. Negative values of the MNF slope are related to the insurgence of fatigue.  $* p < 0.05$, $\smallstar  p < 0.01$}
    \label{fig:mnf}
\end{figure}

The results of MNF elaboration are presented in Fig.\ref{fig:mnf}. All configurations report varying slopes that present both positive and negative values. While the negative values are notably related to fatigue, the positive values of the shift of median frequency are more challenging to interpret. This phenomenon can happen, even though the interpretations are different \cite{Phinyomark2012}: either MNF should increase as the muscle force or load increases, or it should increase as the muscle length or joint angle decreases.
This is harder to assess in the case of this experiment since the movement is cyclical, and the considerations presented refer to very short and muscle-specific experiments. Therefore, the analysis will focus only on the downward shift of MNF.

The rectus femoris results show that the greatest fatigue occurs in Configuration A, without XoSoft, which has the lowest $25^{th}$ percentile, and in Configuration C, which has the lowest mean value.
The biceps femoris shows two configurations, C and D, with a longer negative tail than the others, while Configuration A has most values around zero.
The tensor fasciae latae has its lowest $25^{th}$ percentile in Configuration C and its lowest mean value in Configuration A.
The erector spinae iliocostalis presents an interesting result since it has the lowest $25^{th}$ percentile and the mean value for Configuration A, the only one where the weight of the exoskeleton is not present. This could mean that the muscles affected by the weight of the exoskeleton may be other back muscles since subjects verbally reported feeling tired from the weight of the backpack. Some of the back muscles experiencing fatigue could be the longissimus, responsible for resisting forward flexion; the quadratus lumborum, which may show lower MF under asymmetric or heavy loads; or the trapezius, which contributes to shoulder and upper back stabilization, especially under backpack loads.

The results of Friedman's test show significance for the biceps femoris (p=0.0206) and the erector spinae (p=0.0138), while there is no evidence of significant difference between the configurations for the rectus femoris (p=0.5) and the tensor fasciae latae (p=0.2). 

The Wilcoxon signed rank test was performed on the pairwise distributions of the biceps femoris and the erector spinae since only their Friedman was significant. For the biceps femoris, the configuration that appears more significantly different from the other is Configuration C, with respect to Configuration A (p = 0.0012), B (p = 0.0206), E (p = 0.0522), F (p = 0.0228). Given these results, we can hypothesize that the placement of the anchor points in that configuration introduces a greater stress on the biceps femoris. As for the erector spinae, Configuration A presents a significant difference with Configuration C (p = 0.0152), D (p = 0.040), and F (p = 0.04), and there is also a significant difference between configurations C and E (p = 0.0333).

The Friedman analysis of the effect of accumulated fatigue presents some interesting results: all the muscles analyzed appeared to be affected by this factor, with the exception of the hip flexors. Specifically, the rectus femoris showed no significant effect (p = 0.75), whereas the other muscles exhibited significant or near-significant differences that may help explain the previously observed condition effects, in particular the biceps femoris (p = 0.003), the tensor fasciae latae (p = 0.015), and the spinal erectors (p = 0.0603). These results open the question of whether providing variable assistance, like the one given to the hip flexors by changing the anchor points of the actuator, could be an effective way to prevent the insurgence of muscular fatigue.\\

\begin{figure*}[!ht]
\centering
\subfigure[]{\includegraphics[clip, trim={30 30 0 0},width=0.32\linewidth, height=9cm]{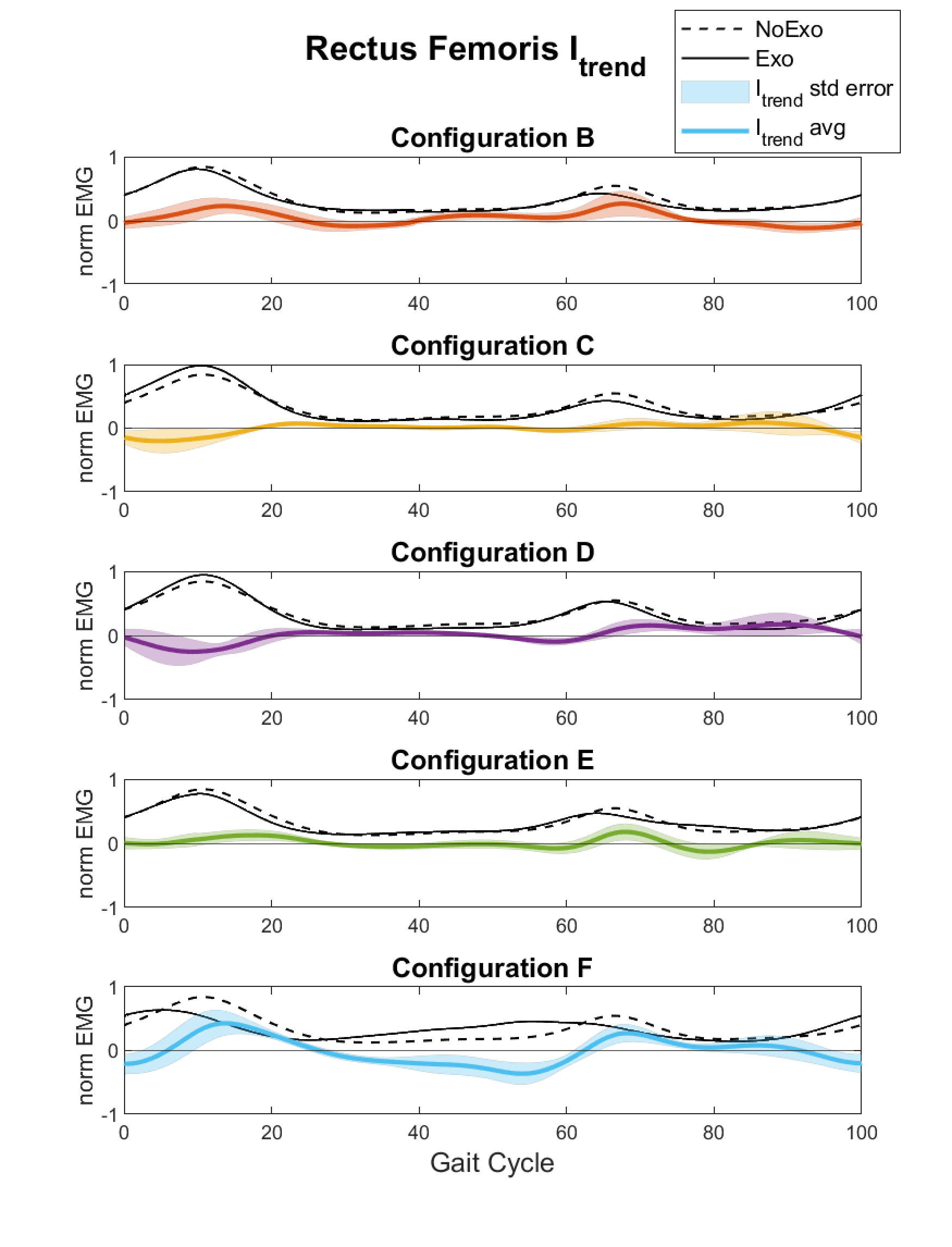}}
\centering
\subfigure[]{\includegraphics[clip, trim={30 30 0 0},width=0.32\linewidth, height=9cm]{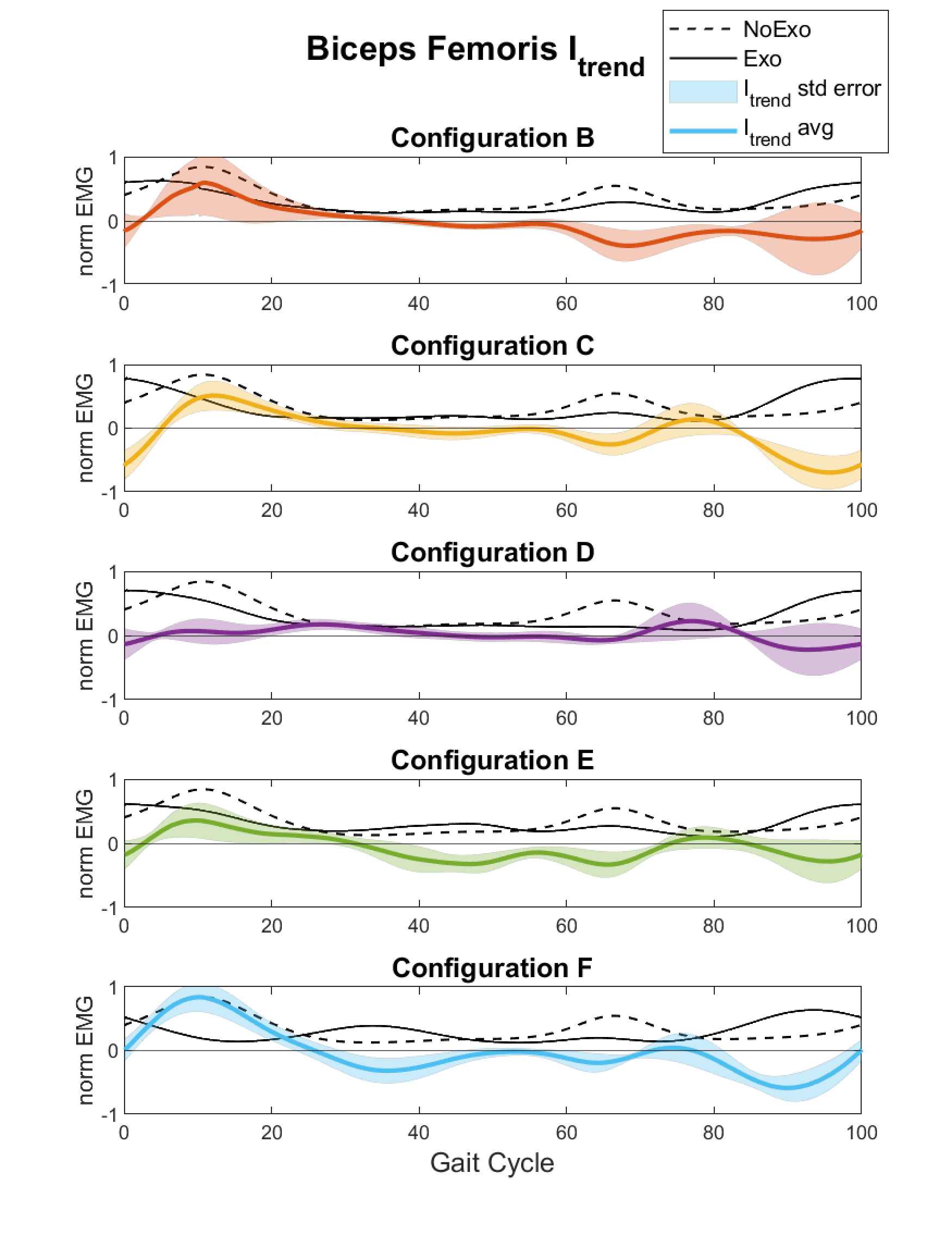}}
\centering
\subfigure[]{\includegraphics[clip, trim={30 30 0 0},width=0.32\linewidth, height=9cm]{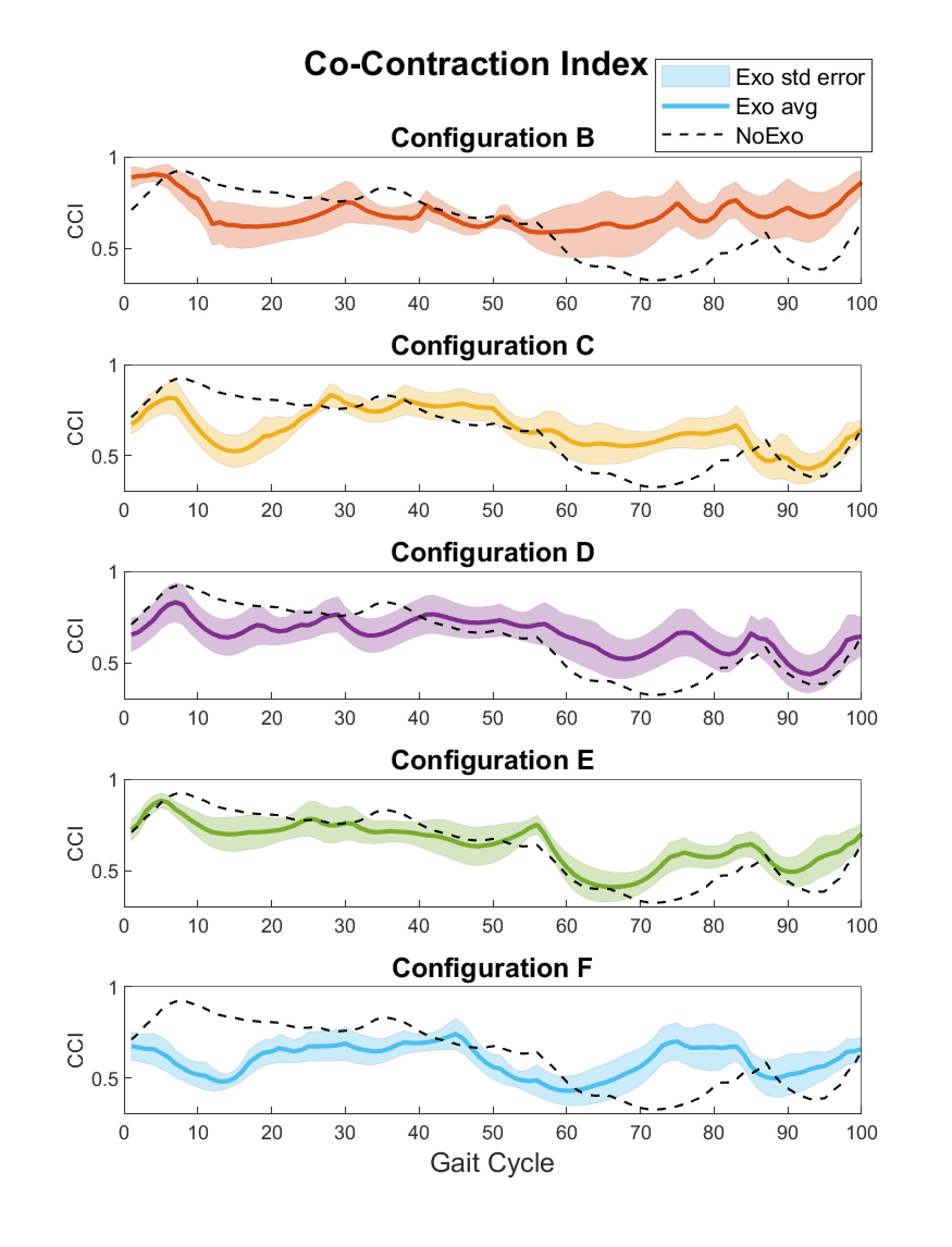}}

\caption{Mean normalized activation of the configuration compared with Configuration A, that is the NoExo condition, with the related standard error and the $I_{trend}$ index of the (a) rectus femoris and (b) biceps femoris. $I_{trend} > 0$ means that the considered configuration is providing assistance if compared to the baseline. (c) Co-contraction index values for the six configurations, for all the subjects involved. All the configurations with the exoskeleton are compared with the baseline state. A CCI value bigger than the baseline one indicates a higher coactivation of the rectus femoris and the biceps femoris in that instant.}
\label{fig:muscular_analysis}
 
\end{figure*}

The average activations of the rectus femoris and biceps femoris with XoSoft are shown in Fig.\ref{fig:muscular_analysis}(a),(b). Rectus femoris activations align with state-of-the-art data (e.g., Winter \cite{winter1991}), while most configurations show a delayed biceps femoris activation compared to baseline.

Along with the activations, the $I_{trend}$ index provides a more immediate visual understanding of the assistive or resistive effect of the configuration considered: when $I_{trend} >0 $, assistance is provided, otherwise, there is a resistive behavior compared to the baseline condition.
To understand the percentage assistance, the Overall Interaction Index (OII) can be used: it is equal to the average activation values \cite{fanti2024}. It evaluates the balance between assistance and resistance and can be calculated by subtracting the resistive instances of $I_{trend}$ from the assistive ones
\begin{equation}
    OII = \sum_i I_{trend >0}(i) - \sum_i |I_{trend <0}(i)|
\end{equation}

Tab.\ref{tab:PV_table} compares the $PV_{90}$, that is the variation of the $90^{th}$ percentile with respect to the No Exo condition, and the $OII$ average values for the different configurations in the rectus femoris and biceps femoris. Both the variations can be interpreted starting from Fig.\ref{fig:muscular_analysis}(a),(b), by looking respectively at the behavior of the peak for the activation with or without the exoskeleton and for the overall assistance provided. 
$PV_{90} > 0$ means that there was an increase in the muscle activation peak, so that the Configuration requires a higher maximum energy. Instead, if $OII > 0$, the period in which the exoskeleton provided assistance is greater than the resistive action period. Hence, there is an overall reduction in the required energy.

\begin{table}
    \centering
    \begin{tabular}{llllll}
         & B & C & D & E & F\\
         \hline\\
        $PV_{90}$ RF [\%] & 17.39 & 0.08 & 0.88 & 10.66 & 8.02\\
        OII RF [\%]& 10.21 & -1.08 & 2.68 & 1.86 & -1.51 \\
        $PV_{90}$ BF [\%]& 14.48 & -0.09 & 3.03 & 8.34 & 5.95\\
        OII BF [\%]& -6.77 & -11.89 & 3.96 & -14.83 & -5.04\\
    \end{tabular}
    \caption{Overall Interaction Index (OII) for the rectus femoris (RF) and the biceps femoris (BF). $OII > 0$ when the exoskeleton provides more assistance than resistance.}
    \label{tab:PV_table}
\end{table}

The peak of the rectus femoris activation results increased for all the configurations; at the same time, the overall interaction index indicates assistive behaviors that benefit this muscle, especially in Configuration B.
For the biceps femoris, all setups except for Configuration C show an increase in the peak activation, while Configuration D is the only one in which the OII suggests overall assistance. From a muscular activation perspective, Configuration D seems promising since the chosen indicators show a very limited increase in the peak activation and an overall assistance for both muscles.
The Wilcoxon sign-rank test was done by comparing the peaks of the single configurations with respect to the baseline for all the subjects. It showed significance only for Configuration E in the biceps femoris (p = 0.0391).\\

The results of the comparison between the co-contraction indexes over the gait cycle are shown in Fig.\ref{fig:muscular_analysis}(c).
Configuration A consistently has a higher CCI in the first half of the gait cycle. This result is partly, since according to the literature \cite{winter1991}, the phase where both the rectus femoris and the biceps femoris have their peak activity is between 90-100\% of the previous gait cycle and the 0-30\% of the current one. This could mean that wearing XoSoft causes a lower coactivation during the stance and a higher coactivation during the swing. This could be explained by the shift in the Biceps Femoris activation that is observed in many of the configurations in Fig.\ref{fig:muscular_analysis}. Another possible explanation is that even in a phase like the swing, where the contact forces with the ground are not present, there is a higher need for muscular activation due to the forces introduced by the elastic bands.

Due to data incompleteness, only the Wilcoxon paired test was performed on the integral of CCI, which showed results towards the significance between Configuration A and B (p=0.0625), and significance between Configuration E and F (p = 0.0469).


\subsection{Metabolic consumption Analysis}

\begin{figure}[!h]
    \centering
    \includegraphics[width=1\linewidth]{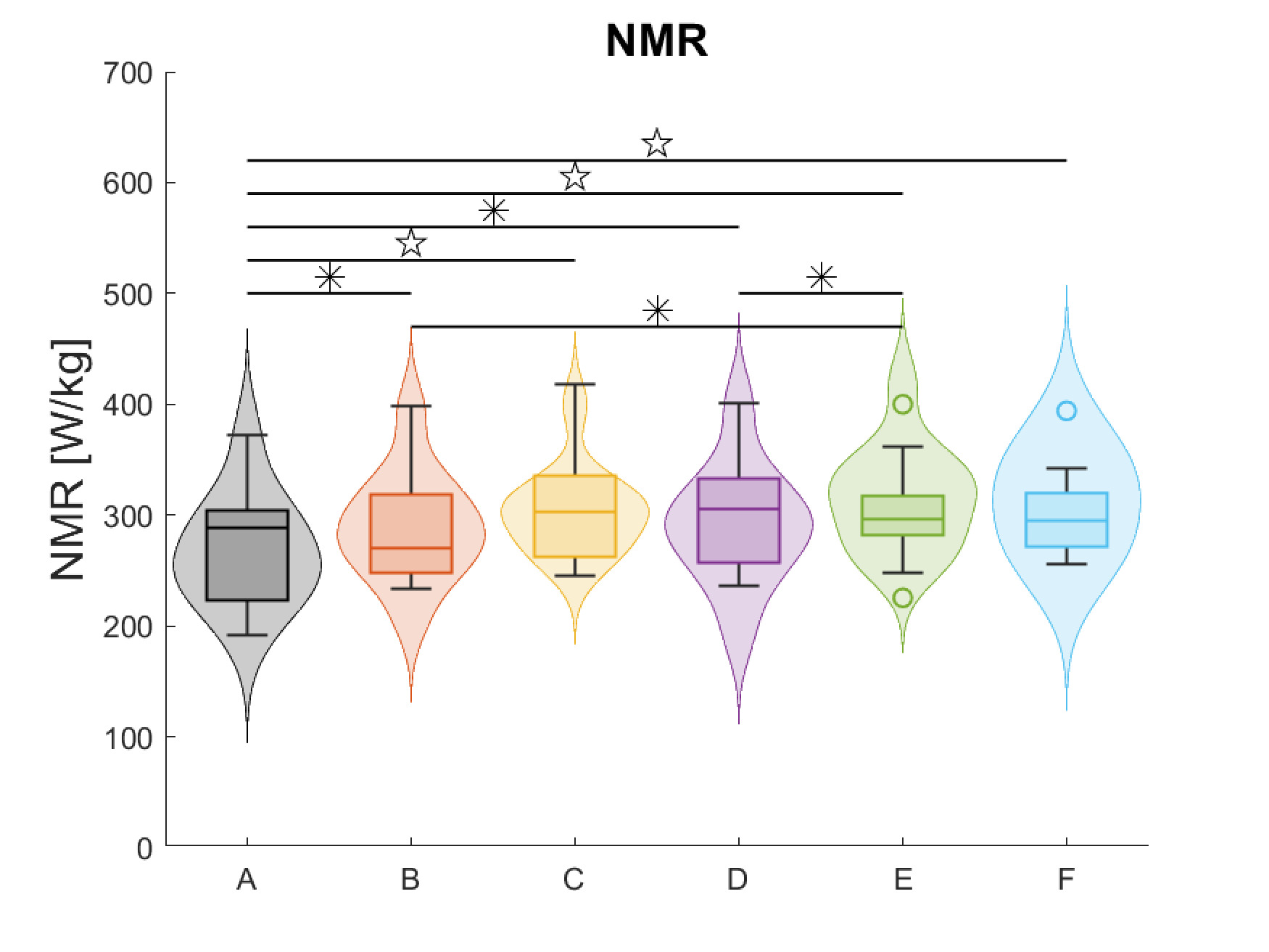}
    \caption{Violin plot of the average NMR for the 11 subjects in each configuration. A higher NMR indicated that the subjects is consuming more energy to perform the same task in another configuration. $* p < 0.05$,$\smallstar  p < 0.01$}
    \label{fig:metabolic_consumption}
\end{figure}

Fig.\ref{fig:metabolic_consumption} shows the impact of all the configurations on the overall group of subjects. The NMR results are lower for Configuration A, the one without XoSoft, which has a mean of 267.3W/kg, while the lowest value for the other configurations is B, with 289.8 W/kg.

Friedman's test revealed statistically significant differences in metabolic consumption across all configurations (p = 0.0024). When Configuration A was excluded, the test still indicated significant differences (p = 0.033), demonstrating that the actuators' routing has a measurable impact on metabolic consumption. 

A Wilcoxon test identified specific pairs with statistically significant differences, as shown in the figure. As expected, Configuration A results are significantly different from all the others, with different p-values: B (p = 0.042), C (p = 0.0029), D (p = 0.042), E (p = 0.0098), F (p = 0.0049). This is also true for Configurations E and B (p = 0.032), E and D (p = 0.032).

Configuration D, in agreement with the muscular analysis results, appears to have the lowest NMR among the configurations with the exosuit. Configurations F and E show the highest and mean median values, respectively, suggesting that these modes require greater energy expenditure from the subjects.

The increase of NMR in all exoskeleton configurations implies that, currently, the device may not offer the support needed to reduce the energy load during physical activity. It may add resistance or conflict with the users' biomechanics, forcing subjects to exert more muscle effort to compensate for the suboptimal assistance.

\begin{figure}[!h]
    \centering
    \includegraphics[clip, trim={60 30 40 0},width=1\linewidth]{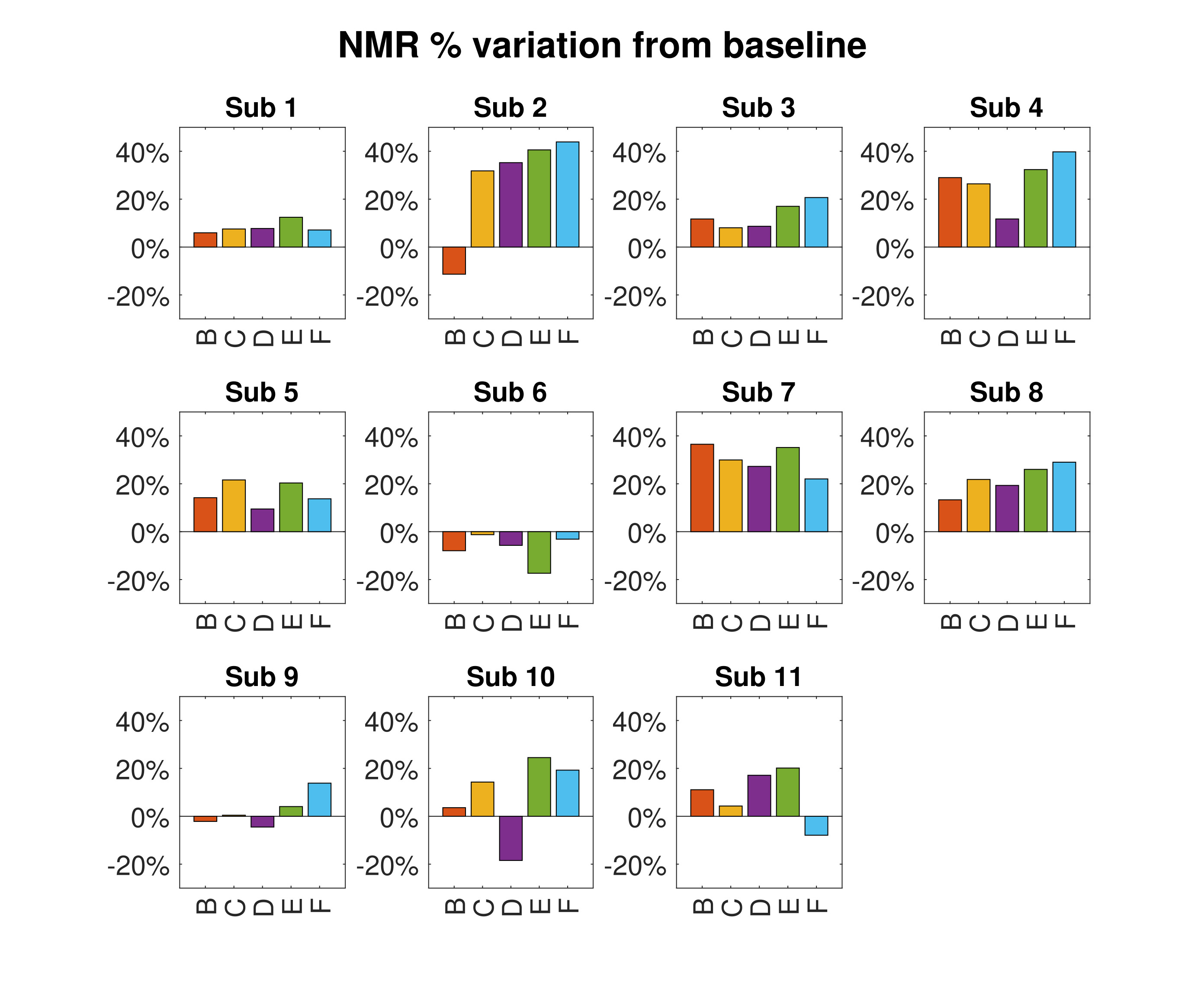}
    \caption{Bar plots representing the difference between the average NMR value of the Configuration and the average ${\text{NMR}}_{{\text{A}}}$, the NMR of Configuration A, normalized by ${\text{NMR}}_{{\text{A}}}$ and expressed as a percentage.}
    \label{fig:metabolic_single_subs}
\end{figure}

The analysis of the single subjects' metabolic consumption in Fig.\ref{fig:metabolic_single_subs} offers a different perspective on the effectiveness of XoSoft in reducing the required energy.
The image shows the difference between the metabolic consumption of the considered Configuration A, normalizing it by ${\text{NMR}}_{{\text{A}}}$ as a percentage.
For individual subjects, it is evident that there is no universally preferred Configuration, and there is great variability in the impact of the exoskeleton on metabolic consumption. Several subjects show a reduction in NMR in some cases. For instance, subject 10 has a decrease of 18.45\% with Configuration D, and subject 6 reduced their metabolic consumption by 17.37\% with Configuration E. Other subjects instead do not appear to benefit from the exosuit and reach increases up to 43.9\%, in the case of subject 2 for Configuration F, or 39.76\% for subject 4 and Configuration F. One cannot generalize the worst configuration either, since even though configurations E and F seem to have the poorest performance, they are still the optimal configurations for subjects 6, 7, and 11.\\

\begin{figure}[!h]
    \centering
    \includegraphics[clip, trim={60 30 40 0},width=1\linewidth]{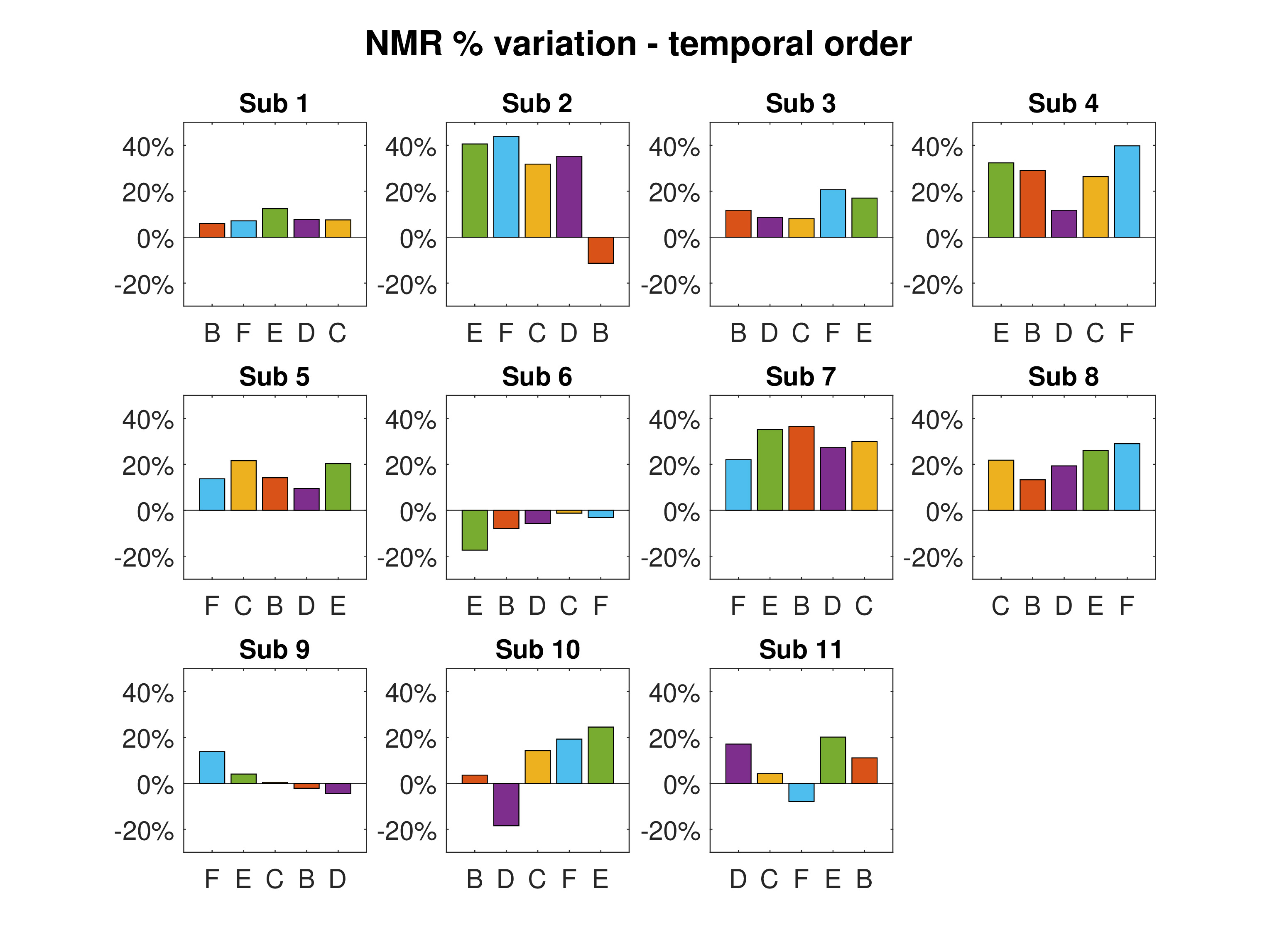}
    \caption{Average NMR results sorted by the order of the configuration presented to the subject. The taller group shows a consistently lower NMR.}
    \label{fig:metabolic_single_subs_time}
\end{figure}

The Friedman analysis of the compound fatigue effect resulted significant (p = 0.0078). The analysis was completed by computing Pearson's correlation coefficient, which revealed a significant positive correlation for subjects 7 and 11 ($\rho$ = 0.89, p = 0.033), a trend for subjects 8 and 10 ($\rho$ = 0.82, p = 0.058), while no significant correlation for the other subjects. The effects of time on metabolic consumption are presented in Fig.\ref{fig:metabolic_single_subs_time}. This more comprehensive analysis, supported by the graphical interpretation of the figure, shows that although cumulative fatigue has a significant impact on some subjects, it does not undermine the previously observed findings. In fact, even in subjects where cumulative fatigue is present, the optimal configuration often does not correspond to the first trial, as observed in subjects 9 and 10.\\

\begin{figure}[!h]
    \centering
    \includegraphics[width=1\linewidth]{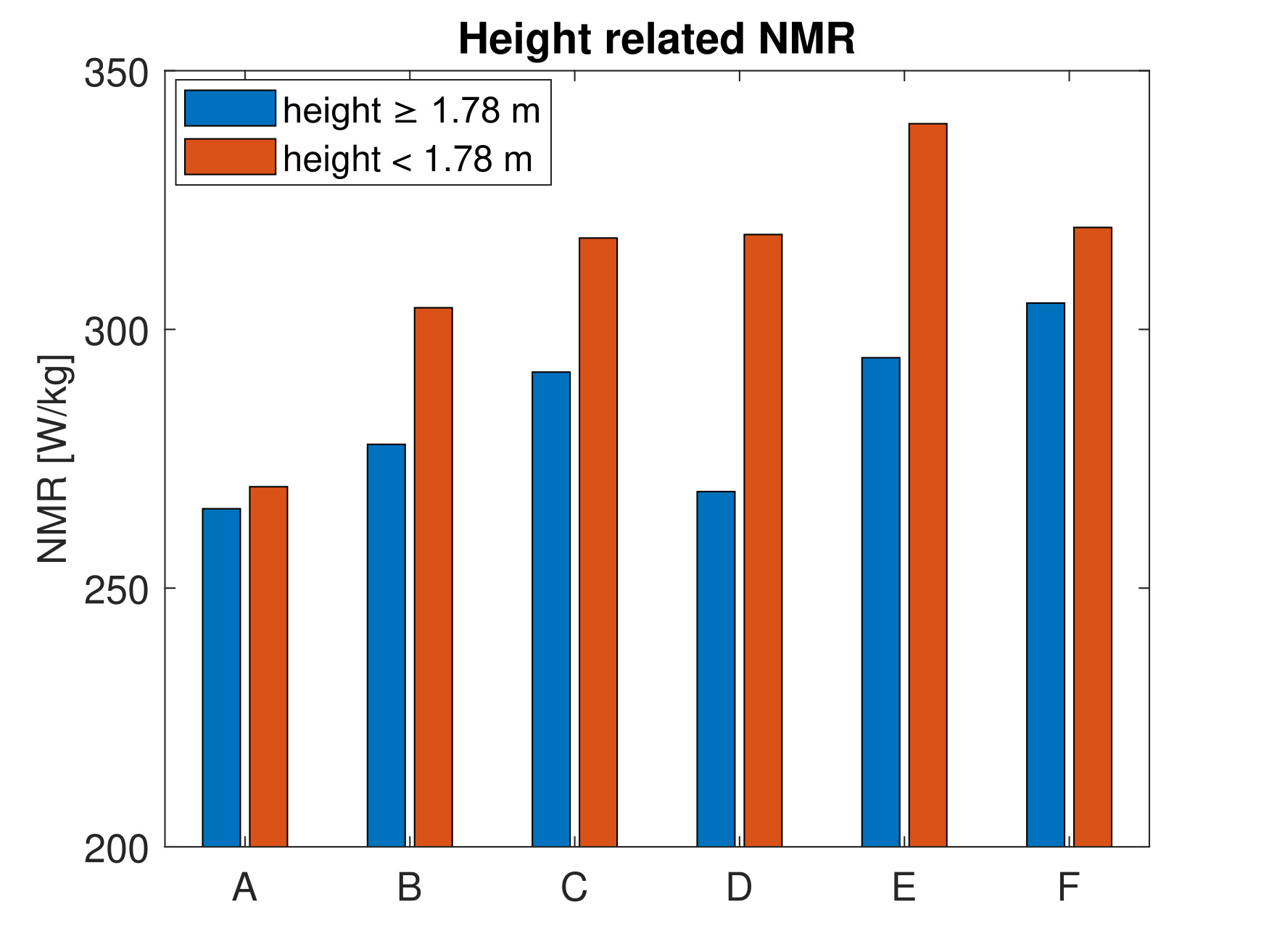}
    \caption{Bar plots representing the average NMR divided by heights groups.}
    \label{fig:metabolic_height}
\end{figure}
An analysis was conducted by comparing performance across height groups, as shown in Fig. \ref{fig:metabolic_height}, to investigate if this could be a factor contributing to a reduction in metabolic consumption using the exoskeleton. The results suggest that taller subjects may benefit more from the assistance. This could be due to the clutch mechanism, which needs quite a long stroke to function and, therefore, can store an adequate amount of elastic energy in individuals with longer limbs.

A Kruskal-Wallis test was performed on the individual configurations to determine whether height had a significant impact. The results were not statistically significant, with the lowest p-value observed for Configuration D (p = 0.1), which is also the only one resulting in an average metabolic reduction compared to the baseline condition.

\section{Discussion}
The results of this study show that different anchor point placements in a quasi-passive exosuit can lead to significant differences in kinematics, muscle activation and fatigue, and metabolic cost.

While some configurations exhibit generally positive trends, such as the benefits of Configuration D on biomechanical similarity to unassisted gait, muscle activation, and metabolic cost, others show undesired effects, like the reduced range of motion in Configuration E. One possible explanation lies in the alignment of anchor points with the natural positioning of muscles because Configuration D follows the path of the hip flexors, which we are assisting with this actuator. At the same time, placing an anchor point behind the knee increases the arm of the force, potentially enhancing efficiency. In contrast, Configurations E and F position the hip anchor point on the posterior side, whereas the hip flexor muscle heads attach anteriorly to the hip joint, possibly leading to less effective force transmission. However, these trends cannot be generalized to determine a universally optimal placement for all subjects. Instead, the findings suggest that a personalized approach may be more effective.

Subject-specific gait characteristics, including those associated with pathological conditions, can be integrated into musculoskeletal simulation models as trajectory and force data or as changes to the musculoskeletal model to inform optimal anchor point placement. These simulations may be used either to estimate the kinematic impact of the exosuit, such as improvements in gait symmetry and joint range of motion, or to evaluate the dynamic interaction between the device and the musculoskeletal system. The latter approach depends on the accuracy of the underlying musculoskeletal model in capturing subject-specific dynamics, which can be particularly challenging in individuals with motor impairments.
For individuals with neurological conditions, such as stroke, models would need to account for the motor adaptations that have occurred post-injury. Given the complexity of these adaptations, a practical first step may be to focus on how anchor points position influence kinematics before attempting detailed metabolic predictions.
This study explored which kinematic metrics could serve as effective descriptors of assistance quality. Among those analyzed, the DTW  distance, particularly at the assisted joint, showed the strongest correlation with metabolic cost ($\rho$ = 0.44, p = 0.0008), approximately double that of the other metrics. While this value indicates a moderate association, it suggests that DTW distance may still serve as a useful proxy for assistance effectiveness, offering a kinematically grounded alternative to direct metabolic cost estimation. 
As highlighted in our previous work \cite{lambranzi2025}, reliance on simplified metabolic models may lead to inaccurate or contradictory predictions compared to experimental outcomes. This further motivates using robust, data-driven kinematic indicators such as DTW distance, especially when working with complex or impaired motor profiles. Looking ahead, DTW distance could serve as a meaningful validation metric in subject-specific simulations, offering a means to compare simulated and experimental joint kinematics. It may also be incorporated into the cost function during model calibration, where minimizing DTW could help align simulated outputs with real-world motion patterns and possibly indicate a reduction in the metabolic consumption. This would support a tighter integration between experimental data and simulation.

Personalized strategies have previously been proposed by Chen \cite{chen2024}, Prasad \cite{Prasad2024}, and Bonab \cite{bonab2024} as a starting point for developing a subject-specific design methodology through simulation, leveraging user-specific anthropometric data and gait trajectory. 
The obtained results confirm the potential of personalized approaches in obtaining the best results for each subject but also show that validating simulation results with real-life experiments remains essential, not only to confirm findings but also to provide new insights into how assistive devices interact with the human body.

Future simulation frameworks of XoSoft should include external factors such as the added mass of the backpack, which is not negligible. Moreover, to better align with experimental outcomes, simulations should aim to estimate total metabolic cost rather than limiting the analysis to specific muscle groups such as the hip flexors. However, a key barrier remains the numerical instability and high computational cost of complex musculoskeletal models, especially when additional external forces are introduced \cite{lambranzi2025}. Overcoming these challenges would enable more accurate, personalized predictions of gait and metabolic impact, potentially reducing the need for extensive physical testing.

Compared with other studies using XoSoft, the results show a reduced and less constant reduction in metabolic consumption compared with \cite{DiNatali2023}, which presents on average a 4.6\% decrease in the setup that assists only hip flexion, 7.6\% while also assisting ankle plantarflexion. This lack of consistency could be caused by the greater variability of subjects present in this work, where not everyone benefits equally from assistance; another possibility is that this is due to the high speed we used in this study, which was designed to induce fatigue in the subjects.

Pathological gait is known to increase metabolic cost compared to healthy individuals, with studies \cite{DETREMBLEUR2003} reporting significantly higher energy expenditure in post-stroke walking.
However, the impact of assistive devices on metabolic cost in impaired populations remains an open question. In neurological patients, improved gait symmetry may reduce energy cost, but this is not guaranteed, as shown by contrasting results for ankle exosuits \cite{bae2018},\cite{awad2020}. For elderly subjects, some studies have shown promising outcomes, both with passive devices \cite{Panizzolo2019} and active systems \cite{Tricomi2024}. Notably, the passive device's performance, similar to XoSoft operating in an unpowered mode with an engaged clutch, suggests that positive effects may be achievable even with low-complexity solutions.

We expect that variability in assistive needs, including the optimal placement of anchor points, will remain a significant factor in individuals with gait impairments. While identifying subject-specific characteristics that influence configuration preferences in healthy individuals is a valuable step, these determinants may not generalize to populations such as older adults or individuals with neurological conditions.

This work highlights a shortcoming of XoSoft Gamma, which is an increase in metabolic consumption for most configurations. A possible solution to address this issue is to create a lighter prototype that provides the same support in terms of force. This aligns with findings from load-carrying studies with back exoskeletons \cite{fanti2024_1}, which demonstrated that metabolic benefits from exoskeletons appear only when external loads exceed \(\sim 5\,\mathrm{kg}\); below this threshold the added device mass can negate any assistance benefit.
In addition to reducing weight, adjusting assistance through a variable control strategy may be effective. As shown by the behavior of the rectus femoris, providing variable assistance could improve efficiency. Working on the design and the control system together will likely lead to better metabolic outcomes in future versions of the exosuit. The current controller uses a finite state machine with a latency \cite{dinatali2020}, this delay is compensated for by an offset on the gait cycle percentage, but it does not adapt to variations in walking cadence. Additionally, the control timing is not currently adaptable. These limitations will be addressed in future developments of the control system, with the goal of improving responsiveness and adaptability to user-specific gait dynamics.

This study presents some limitations, including the small and homogeneous sample population. Furthermore, the experiment duration was at the lower limit for a reliable metabolic consumption assessment, yet still demanding for the participants. Future work should consider these limitations, with a possible solution in longer experiments with fewer combinations on more subjects.

While these findings provide useful information about the effects of anchor point placement in healthy individuals, we will need to further research how this affects subjects with pathological gait, as their neuromuscular and biomechanical adaptations may lead to different responses to exosuit assistance.

Variations in the considered metrics demonstrate that anchor points positioning is crucial in designing and fitting a device that meets the user's needs and provides effective assistance.
This leads to the key questions of how to identify the optimal configuration for each individual and how to develop simulations capable of addressing this challenge, which should be considered in future work on the topic.

\section{Conclusion}
This paper presents the biomechanical and energetic analysis of the effect that different positions of the anchor points of a lower limb exosuit can have, addressing the lack of experimental validation in the literature on anchor points placement.

This study demonstrates that the points where the actuators are attached in a quasi-passive exosuit significantly affect kinematics, muscle activation and fatigue, and metabolic cost. While specific configurations showed benefits in reducing muscle activation and metabolic expenditure, others restricted the range of motion or produced an increase in energy expenditure. However, no universally optimal configuration was identified, highlighting the need for a personalized approach to exosuit design.

Future work should focus on validating the approach on clinical populations and exploring long-term adaptation effects to enhance the usability and effectiveness of exosuits in rehabilitation and daily assistance.

\bibliographystyle{ieeetr} 
\bibliography{bibliography}

\end{document}